\documentstyle[astrobib,times,psfig]{mn_daw}

\pssilent
\def\imgfig{

\begin{figure*}

  \parbox{0.49\textwidth}{
    \psfig{figure=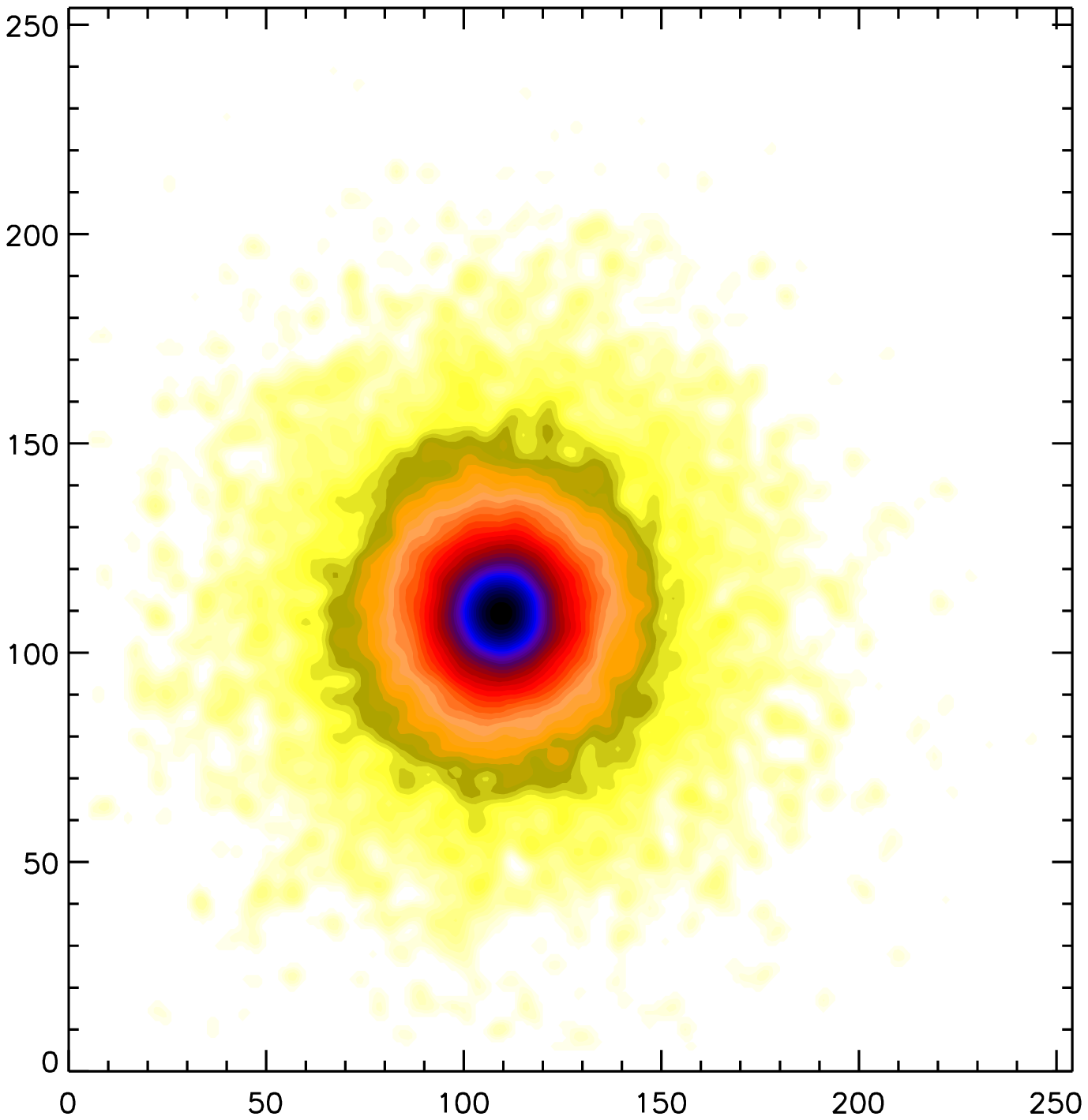,angle=0,width=0.425\textwidth,height=0.425\textwidth}
    \centerline{(a)}
  }
  \parbox{0.49\textwidth}{
    \psfig{figure=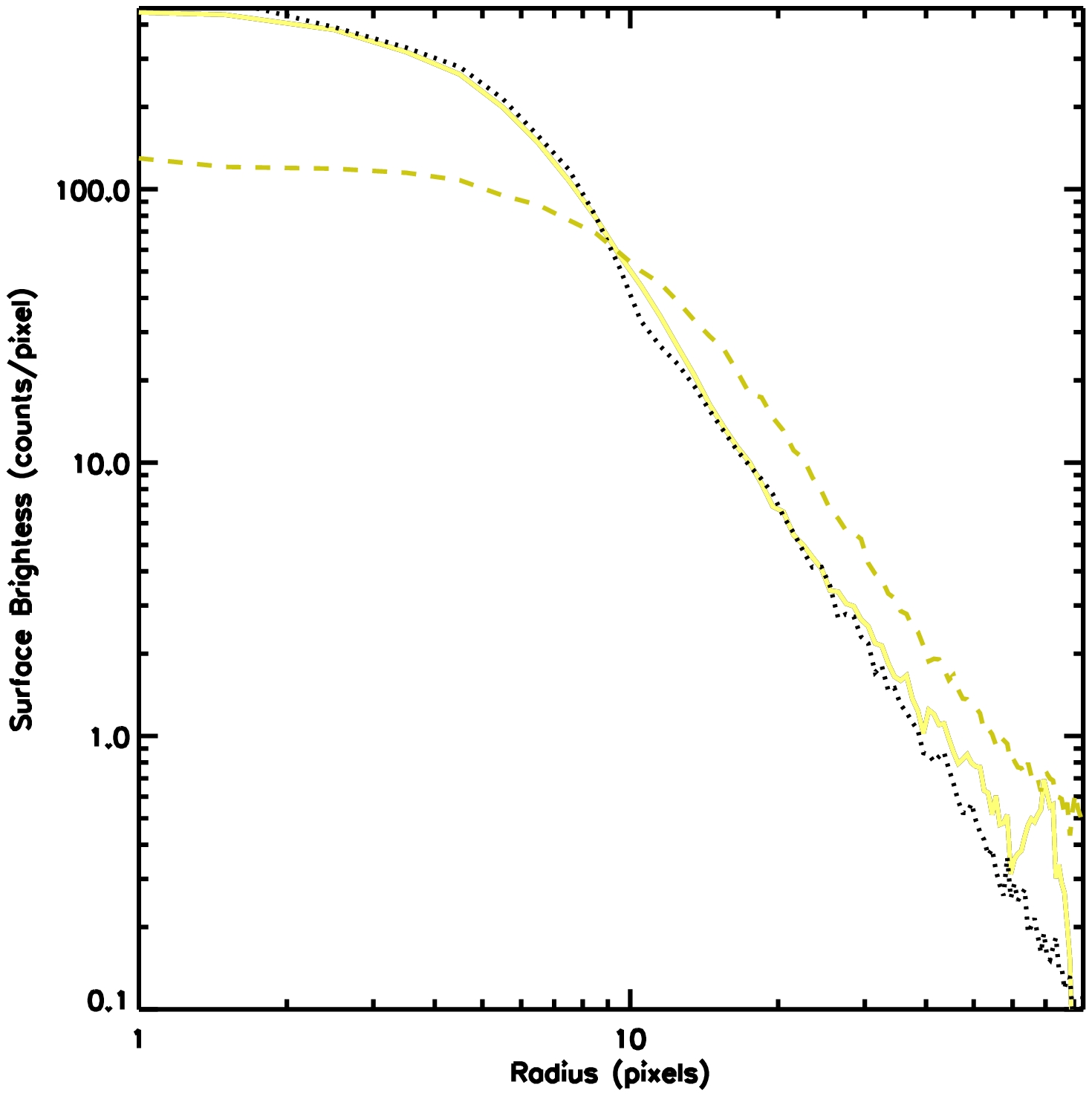,angle=0,width=0.425\textwidth,height=0.425\textwidth}
    \centerline{(d)}
  } 
  \parbox{0.49\textwidth}{
    \psfig{figure=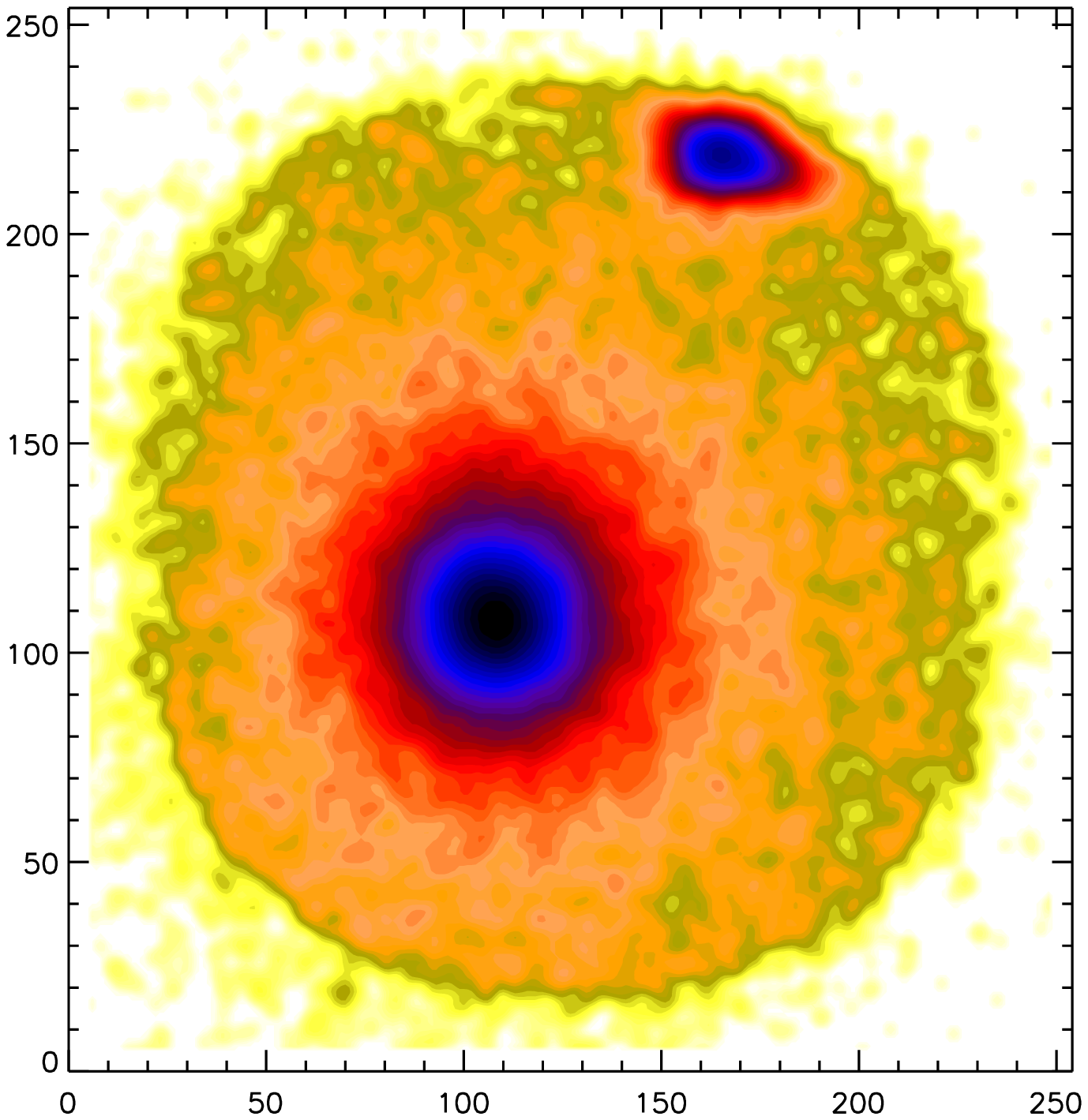,angle=0,width=0.425\textwidth,height=0.425\textwidth}
    \centerline{(b)}
  }
  \parbox{0.49\textwidth}{
    \psfig{figure=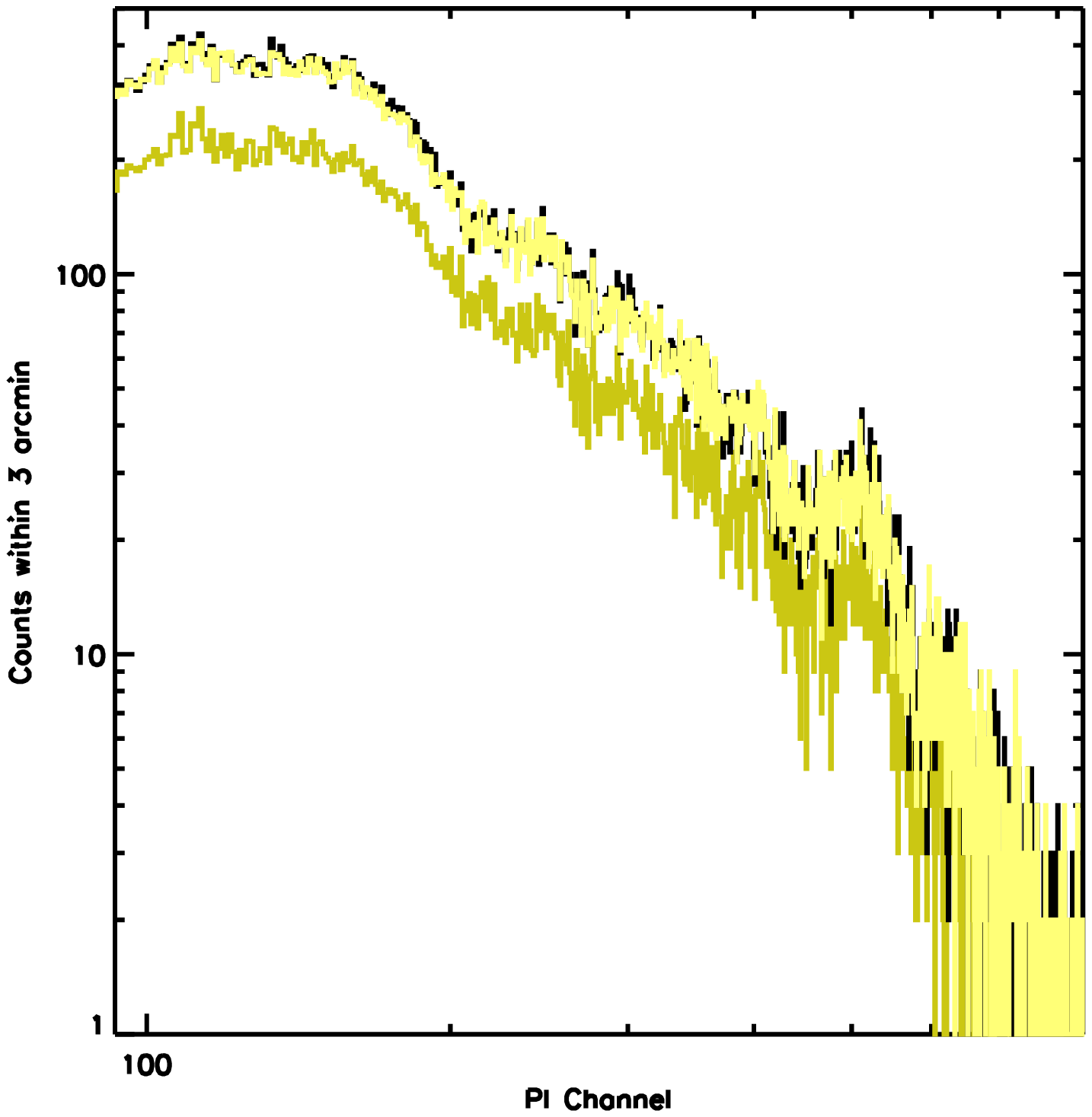,angle=0,width=0.425\textwidth,height=0.425\textwidth}
    \centerline{(e)}
  } 
  \parbox{0.49\textwidth}{
    \psfig{figure=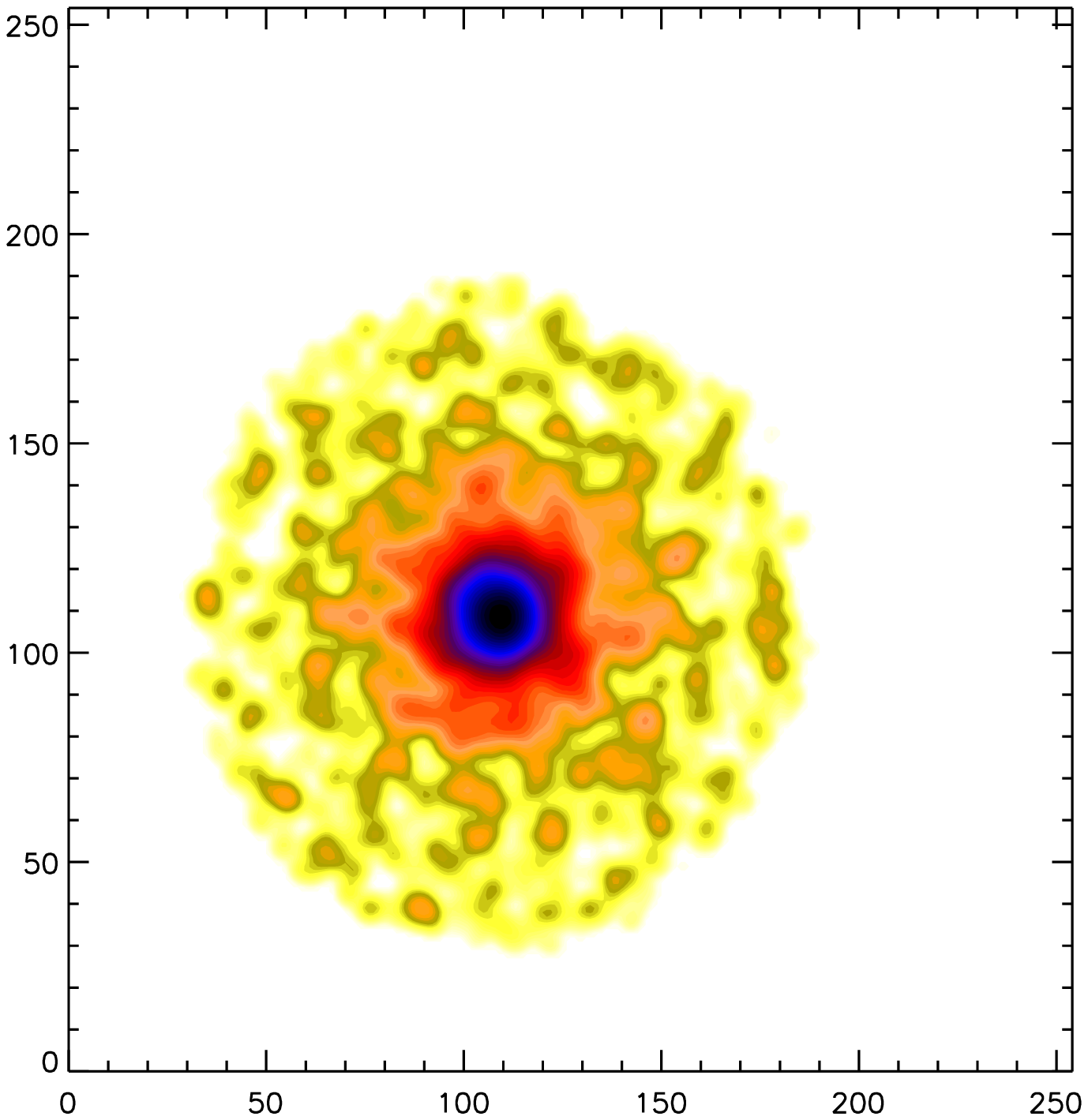,angle=0,width=0.425\textwidth,height=0.425\textwidth}
    \centerline{(c)}
  }
  \parbox{0.49\textwidth}{

	\caption{\label{figure:imgfig}} This figure compares the
	spatial and spectral characteristics of the original,
	convolved and deconvolved datasets.  In (a) we show the image
	of the ({\em TEST-GCF-30K}) events before spatial convolution
	with the PSF and the addition of background events. The
	inclusion of these effects produces in the image shown in (b)
	-- the calibration source in the top-right of the image comes
	from the blank-sky background data. The image shown in (c) is
	that obtained after the spectral-image deconvolution analysis
	-- it is truncated at $20\arcmin$ radius from the centre of
	the field (before the deconvolution) to eliminate the
	calibration source in (b). (These images have a pixel scale of
	approximately $0.25\arcmin$, have been smoothed with a
	Gaussian of 2-pixel width, and are displayed in logarithmic
	intensity). Plot (d) shows the ($1-9\keV$) azimuthal profiles
	of the deconvolved (solid line), convolved (dashed), and the
	original data (dotted line) (from non-smoothed data). The
	upturn in the convolved and deconvolved profiles is due to the
	background, which is not included in the original data. In the
	final panel (e), we compare the spectral PI distribution of
	events extracted from within $3\arcmin$ of the peak
	position. The lower line is the spectrum from the convolved
	dataset. The distribution of the original and deconvolved
	events, from the same aperture, are plotted above as a black
	line and lighter line respectively. This plot shows, firstly,
	that the effect of the PSF is significant -- the spectral
	distribution of the convolved events from the central
	$3\arcmin$ are lower in normalisation (emissivity) and also
	exhibits a different slope (temperature). Secondly, the plot
	shows that the spectral-imaging reassignment recovers the
	characteristics of the original data at {\em full\/} spectral
	resolution.)  }

\end{figure*}

}

\def\psffig{

\begin{figure*}

  \parbox{0.49\textwidth}{
    \psfig{figure=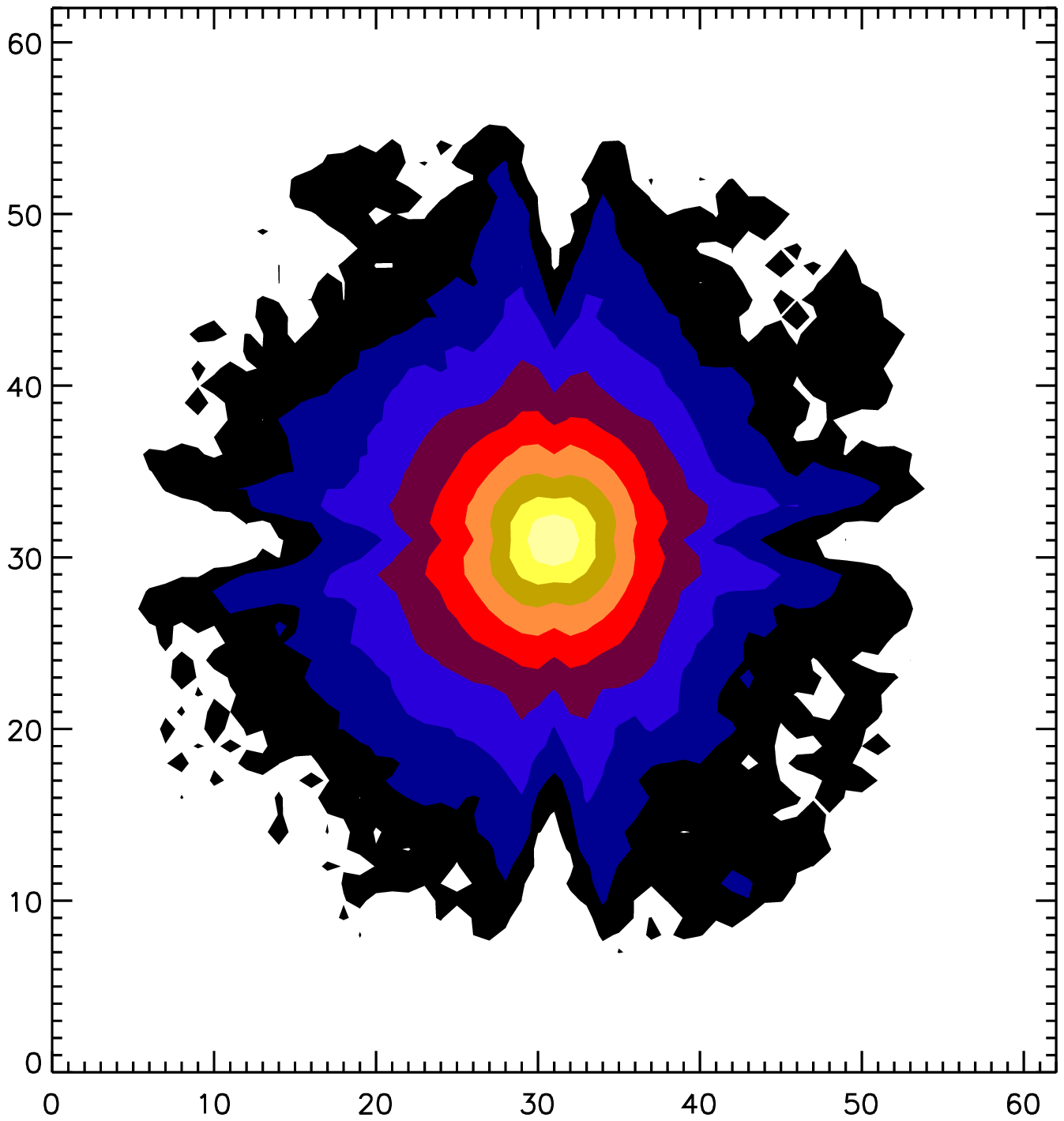,angle=0,width=0.425\textwidth,height=0.425\textwidth}
    \centerline{(a)}
  }
  \parbox{0.49\textwidth}{
    \psfig{figure=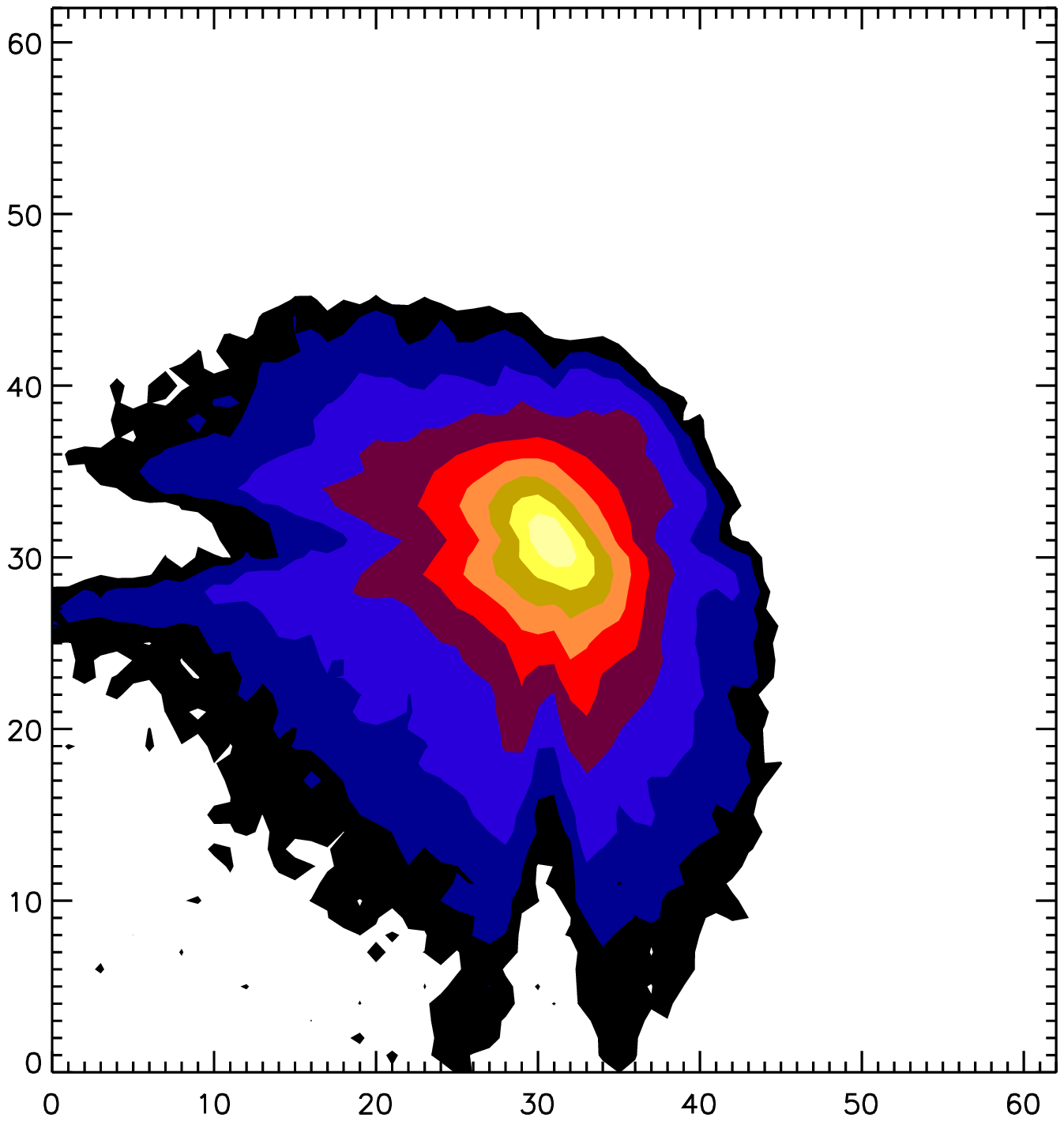,angle=0,width=0.425\textwidth,height=0.425\textwidth}
    \centerline{(b)}
  }
  \parbox{0.99\textwidth}{ \caption{\label{figure:psffig}} In plots
	(a) and (b) we show (in logarithmic intensity) the combined
	XRT and GIS PSF ($1-2\keV$) at two extreme positions:
	$1.8\arcmin$ off-axis with a position-angle of $5\deg$ and
	$13\arcmin$ off-axis with a position-angle $40\deg$,
	respectively.  This highlights the change of the PSF with
	position in the detector image.
	}

\end{figure*}

}

\def\testtak{
\begin{figure}
	\parbox{0.49\textwidth}{
		\psfig{figure=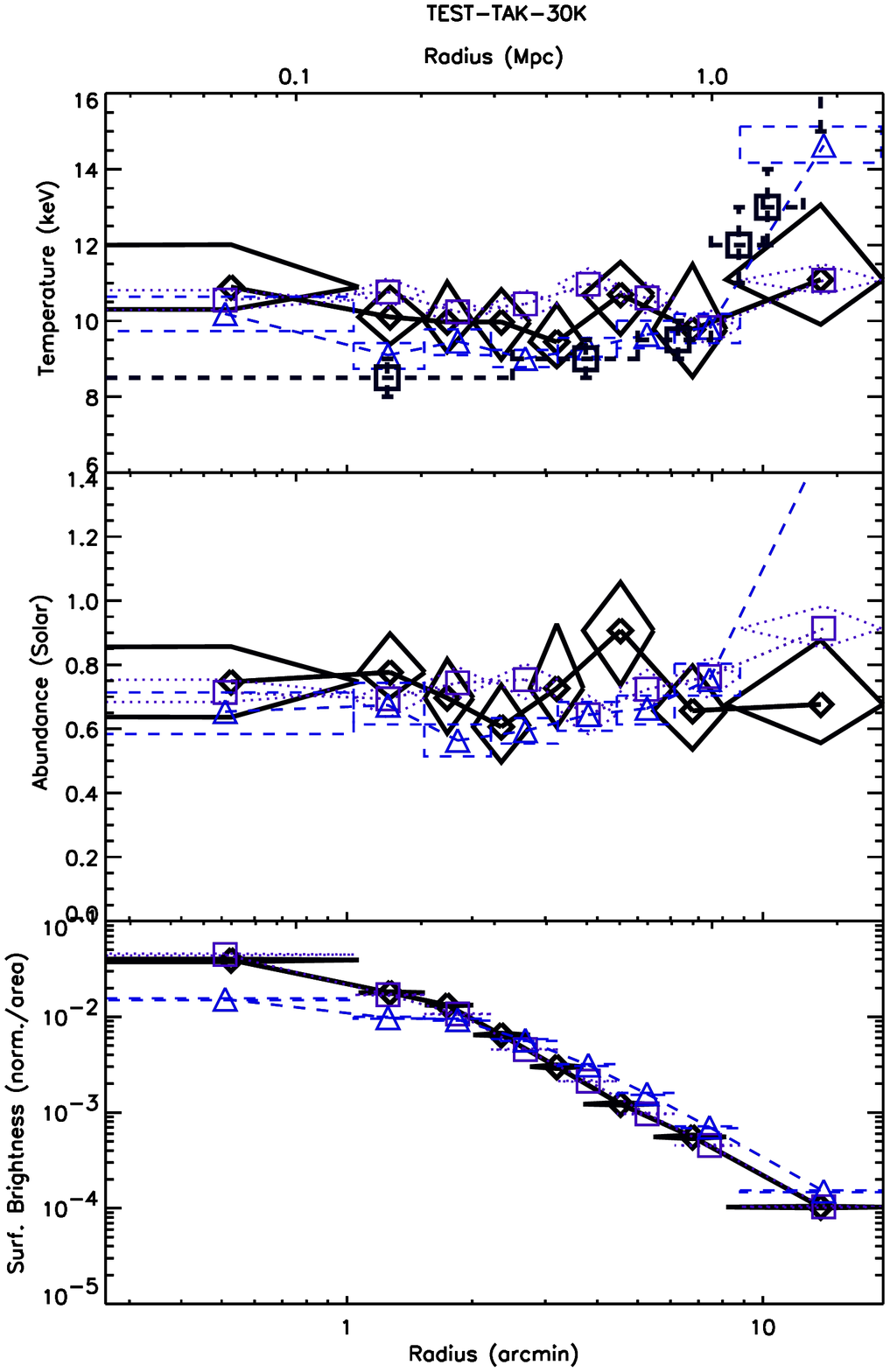,angle=0,width=0.49\textwidth,height=0.79\textheight}
		} \parbox{0.49\textwidth}{
		\hfill\parbox{0.49\textwidth}{
		\captionfont\caption{\label{figure:testtak} This
		figure shows the effect of the PSF on isothermal
		cluster data (the underlying original isothermal data
		is shown by the square symbols with the dotted
		line). Using a model similar to that of Takahashi
		\etal\ (which is shown in the temperature profile
		panel only -- by the square symbols with dashed-line
		error bars) we also obtain the radially increasing
		temperature profile in the uncorrected convolved data
		(dashed line, triangle symbols). This last profile
		(solid line with diamond symbols) shows the recovery
		of the isothermal profile from the convolved data
		using our \SID\ procedure. (See
		Fig.~\ref{figure:teststd} for other plot details.)}}}

\end{figure}
}

\def\teststd{
\begin{figure}
	\parbox{0.49\textwidth}{
		\psfig{figure=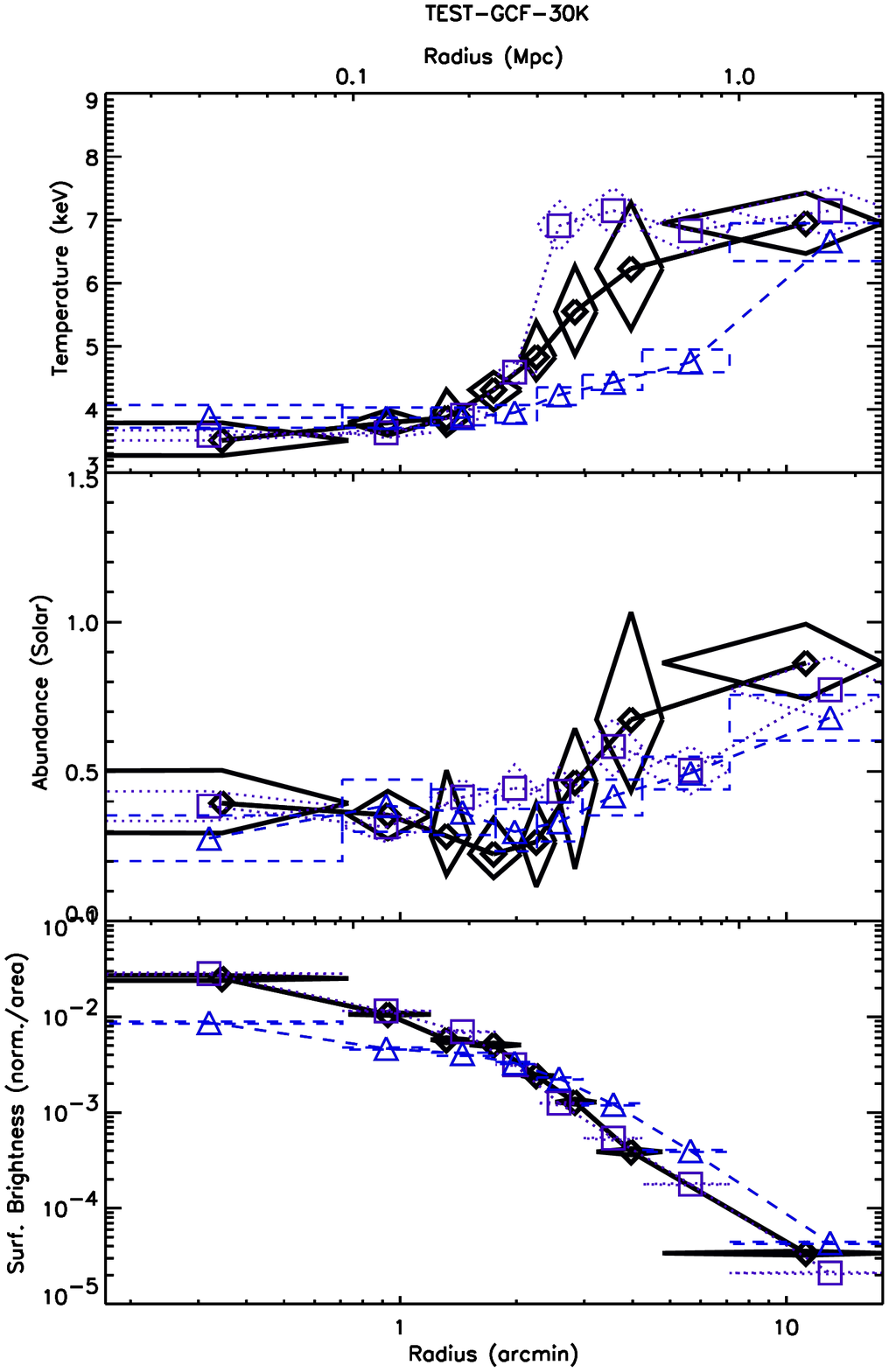,angle=0,width=0.49\textwidth,height=0.79\textheight}
		} \parbox{0.49\textwidth}{
		\hfill\parbox{0.49\textwidth}{
		\captionfont\caption{\label{figure:teststd} In this
		figure we plot the radial variations of electron
		temperature, metallicity (fraction of Solar), and
		model normalisation (\ie\/ emission measure:
		$10^{14}/(4\pi D_L^2) \int n_{\rm e} n_{\rm H} dV$)
		for a single thermal (MEKAL) component fit to the
		standard model ({\em TEST-GCF-30K}; see
		Section~\ref{section:results}). The results for the
		deconvolved data are shown by the solid line, the
		convolved data by the dashed line, and the original
		data by the dotted line. The error bars on the
		temperature, abundance and normalisation are one
		standard deviation, while those on the x-axis data
		show the radial extent.}}}

\end{figure}
}

\def\testtx{
\begin{figure*}
	\parbox{0.49\textwidth}{
		\psfig{figure=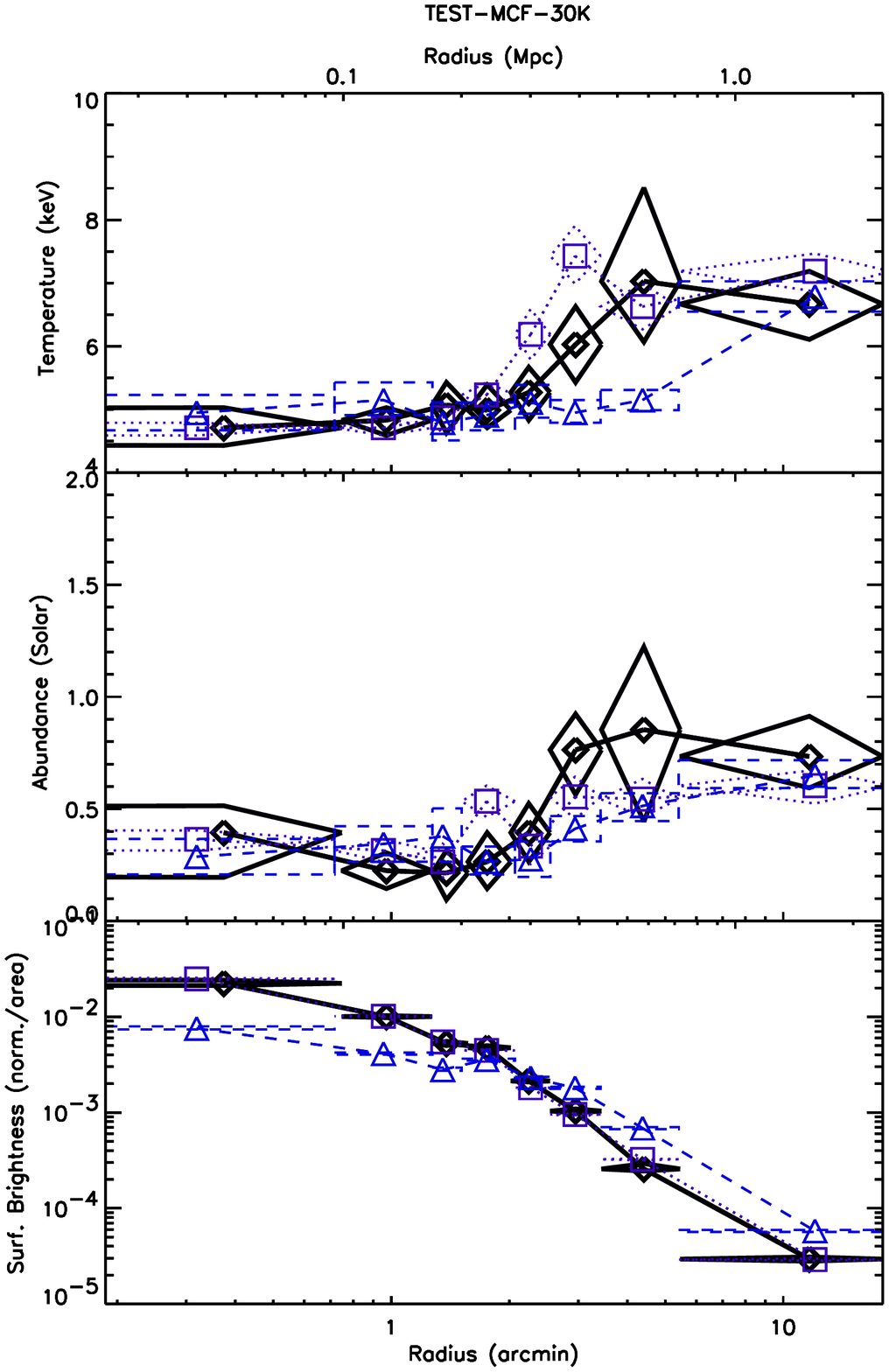,angle=0,width=0.49\textwidth,height=0.79\textheight} \centerline{(a)} }
	\parbox{0.49\textwidth}{
		\psfig{figure=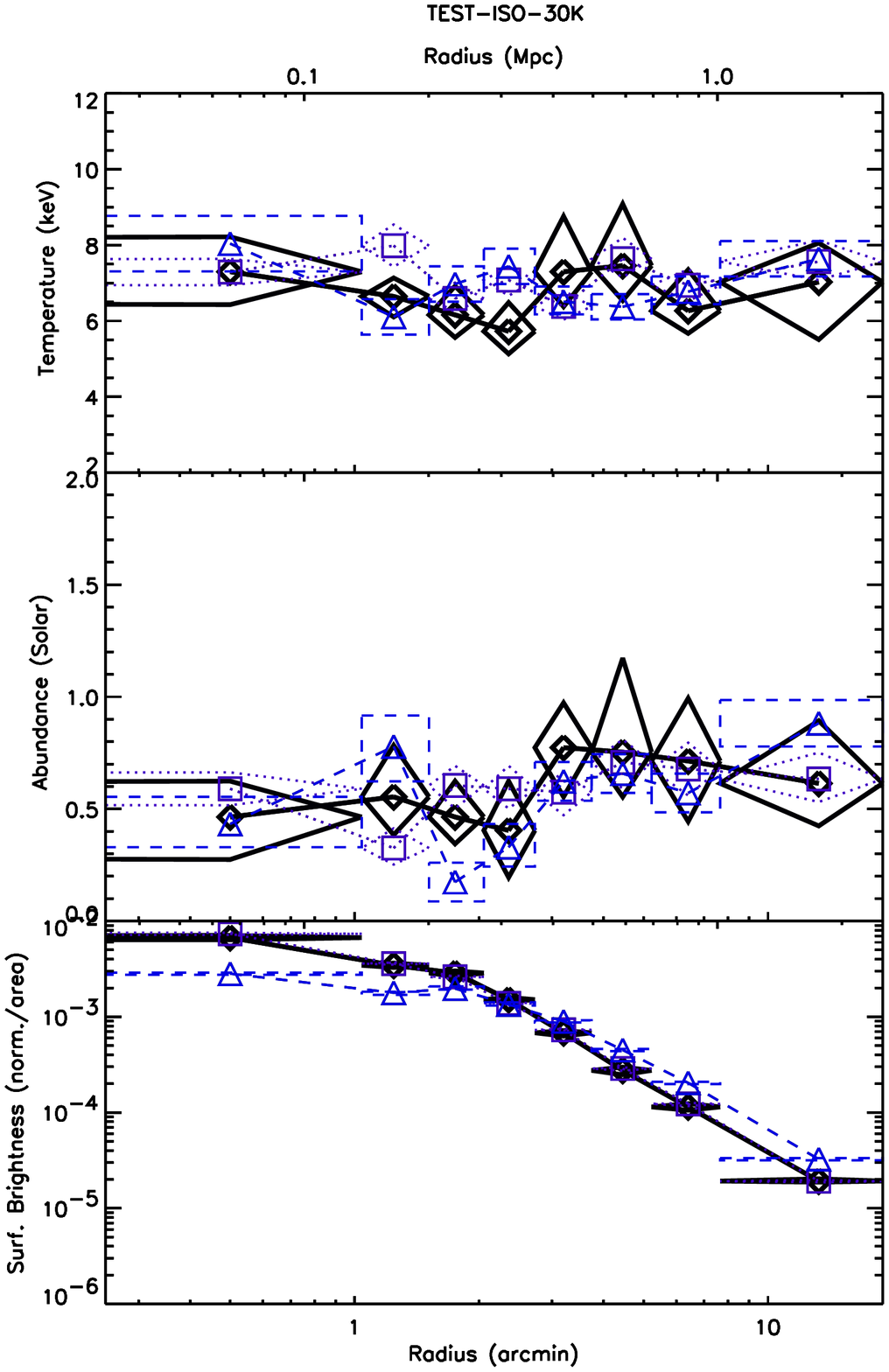,angle=0,width=0.49\textwidth,height=0.79\textheight}
		\centerline{(b)} } \parbox{0.99\textwidth}{
		\hfill\parbox{0.99\textwidth}{
		\captionfont\caption{\label{figure:testtx} This figure
		summarises the tests of differing temperature
		profiles, as follows: (a) moderate core temperature
		drop, (b) and isothermal temperature profile, and (c)
		and temperature decline with radius.  The line styles
		\etc\/ are the same as those defined in
		Fig.~\ref{figure:teststd}.}}}
\end{figure*}
}
\def\testtxcontd{
\begin{figure}
	\parbox{0.49\textwidth}{
		\psfig{figure=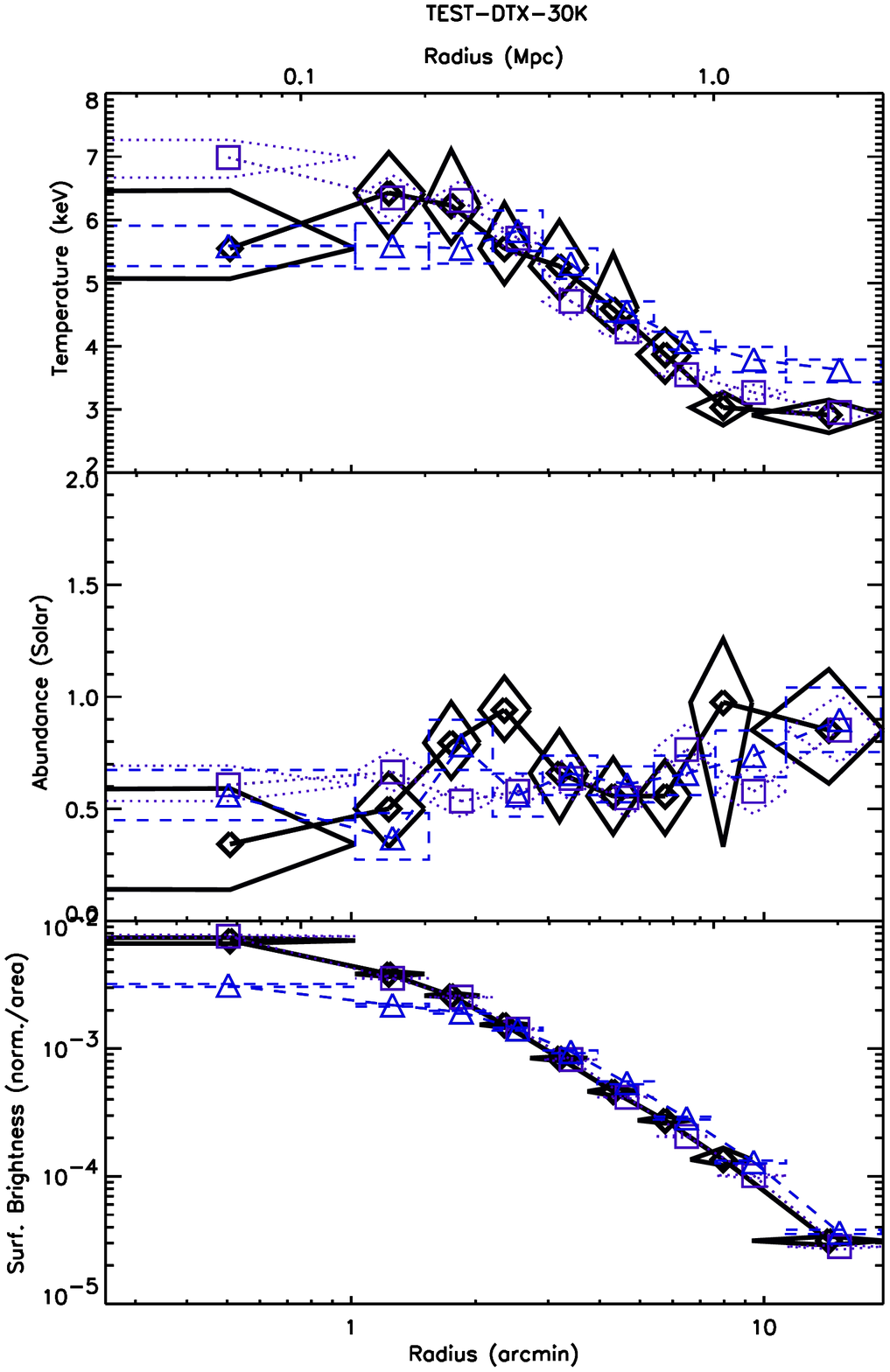,angle=0,width=0.49\textwidth,height=0.79\textheight} \centerline{(c)} }
	\parbox{0.49\textwidth}{
		\hfill\parbox{0.49\textwidth}{
		\addtocounter{figure}{-1}
		\captionfont\caption{
		\label{figure:testtxsupp} -- continued.}}}
\end{figure}
}

\def\testbgd{
\begin{figure}
	\parbox{0.49\textwidth}{
		\psfig{figure=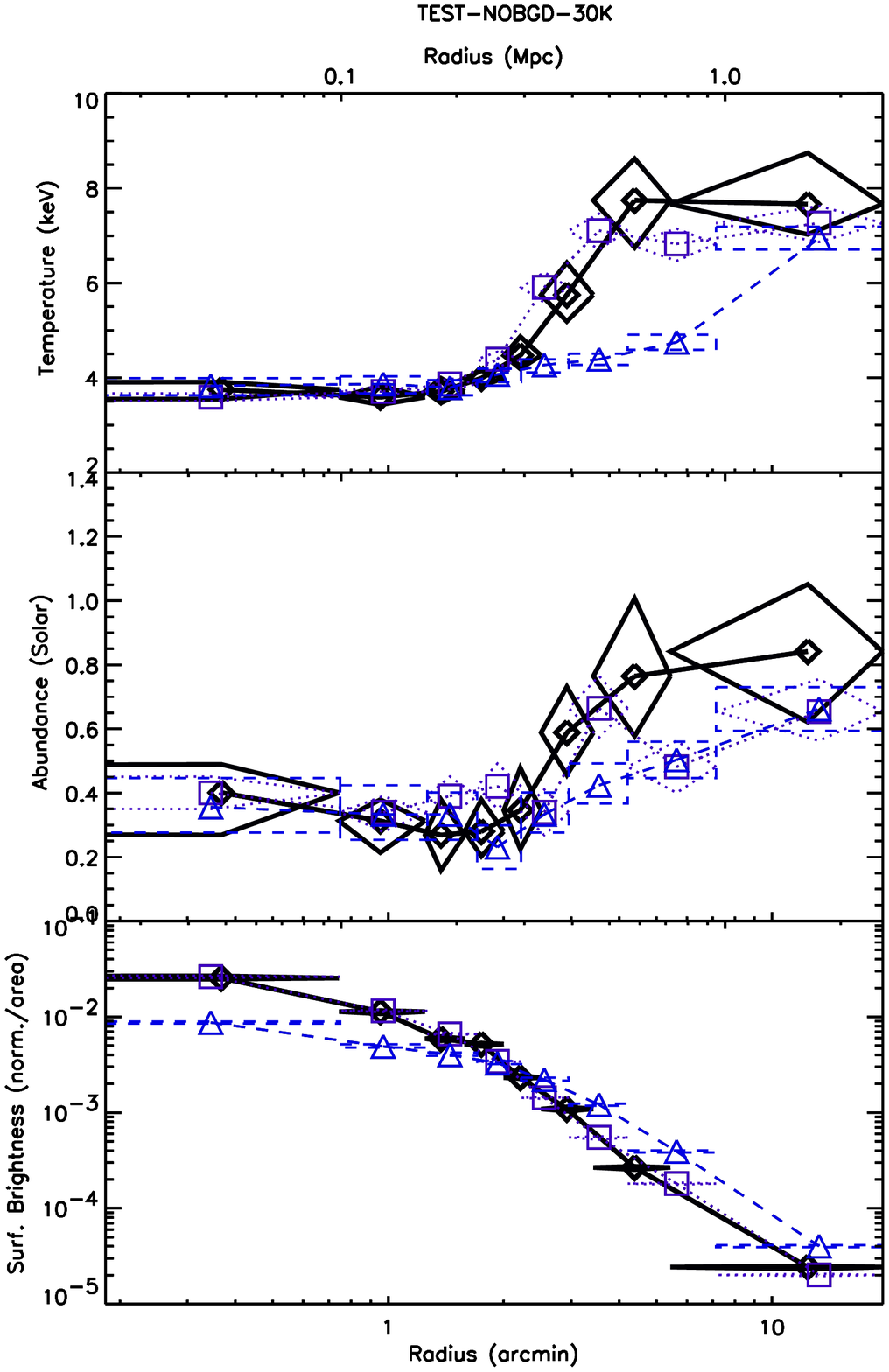,angle=0,width=0.49\textwidth,height=0.79\textheight} }
	\parbox{0.49\textwidth}{
		\hfill\parbox{0.49\textwidth}{
		\captionfont\caption{\label{figure:testbgd} This figure
		summarises the test where no background is included, and
		therefore shows that performance of the \SID\/ method in
		optimal circumstances.}}}
\end{figure}
}

\def\testexp{
\begin{figure*}
	\parbox{0.49\textwidth}{
		\psfig{figure=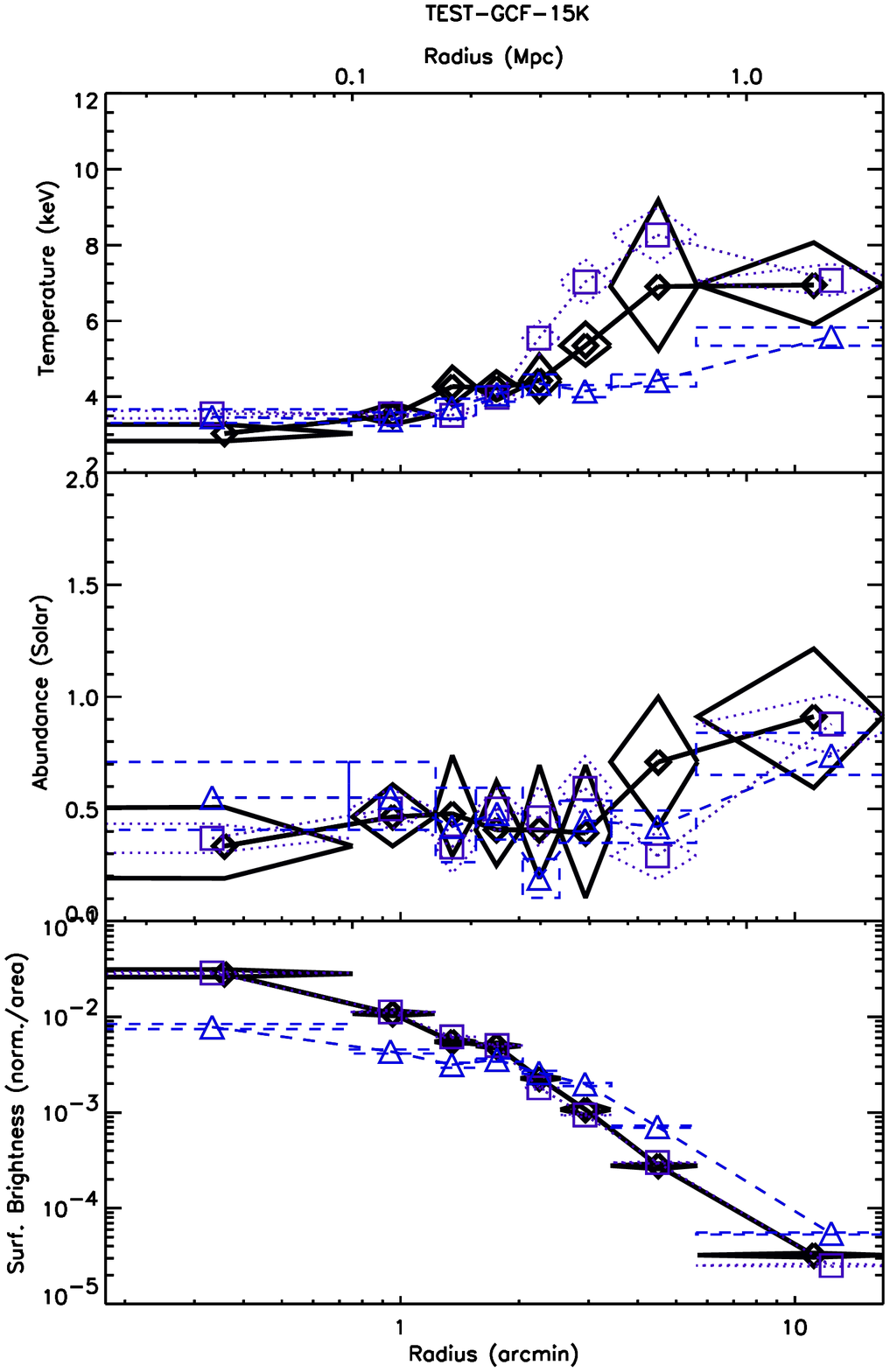,angle=0,width=0.49\textwidth,height=0.79\textheight} \centerline{(a)} }
	\parbox{0.49\textwidth}{
		\psfig{figure=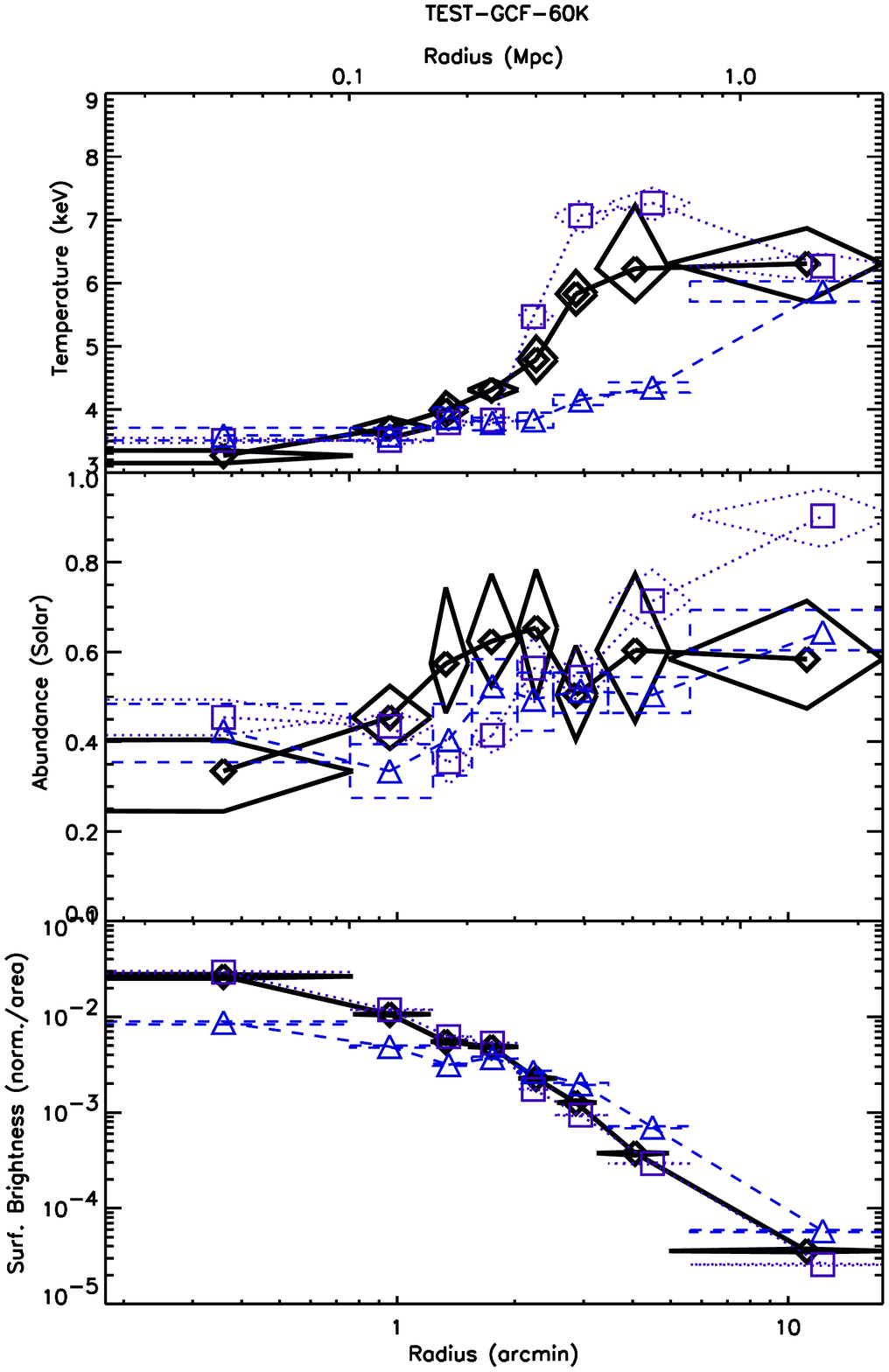,angle=0,width=0.49\textwidth,height=0.79\textheight}
		\centerline{(b)} } \parbox{0.99\textwidth}{
		\hfill\parbox{0.99\textwidth}{
		\captionfont\caption{\label{figure:testexp} This
		figure summarises the test which investigates the
		effect of signal-to-noise through different exposures
		[(a): 15 and (b): $60\ksec$, \ie\/ a factor of
		one-half and twice the standard test's exposure].}}}
\end{figure*}
}

\def\testnit{
\begin{figure*}
	\parbox{0.49\textwidth}{
		\psfig{figure=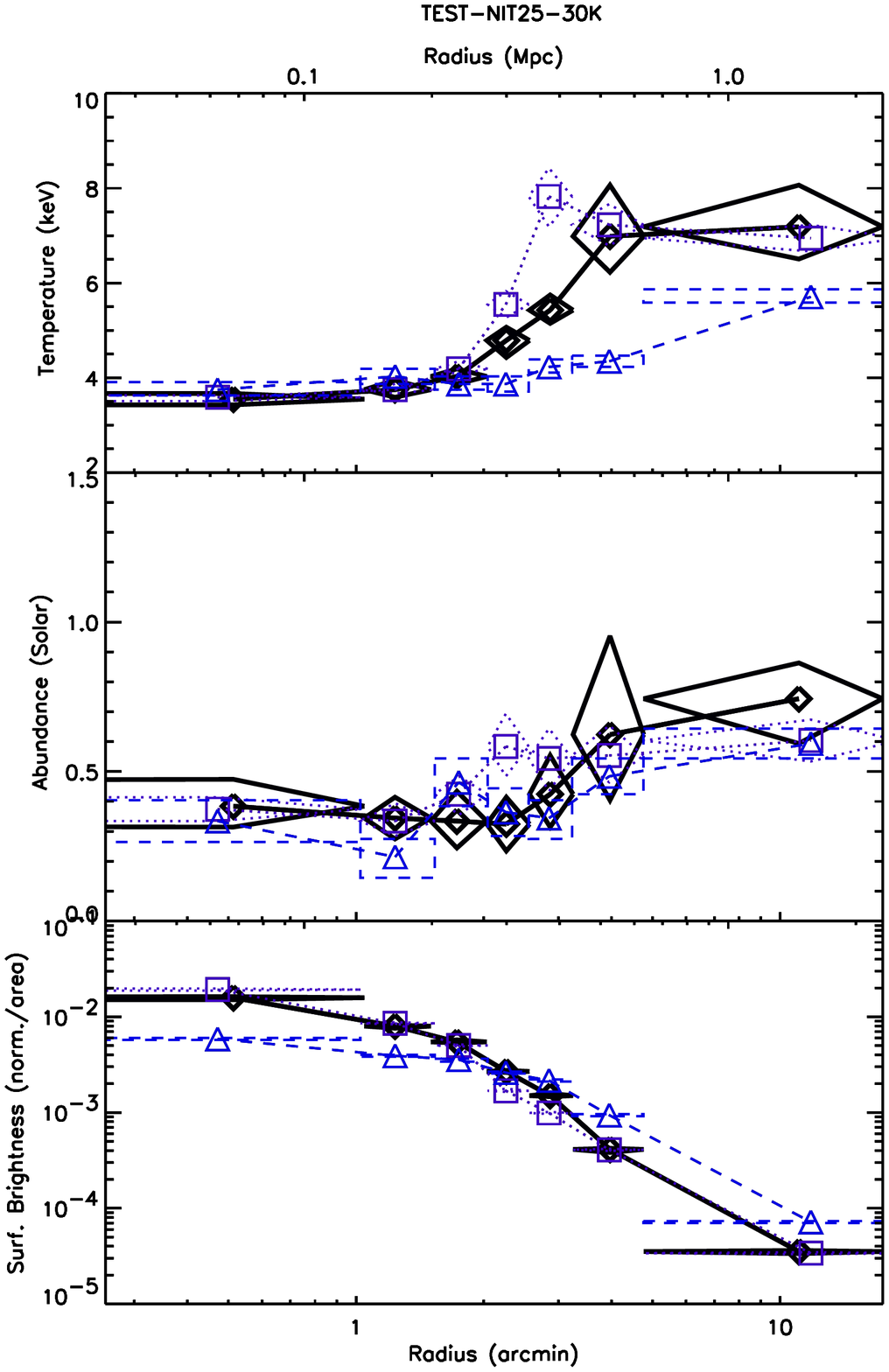,angle=0,width=0.49\textwidth,height=0.79\textheight} \centerline{(a)} }
	\parbox{0.49\textwidth}{
		\psfig{figure=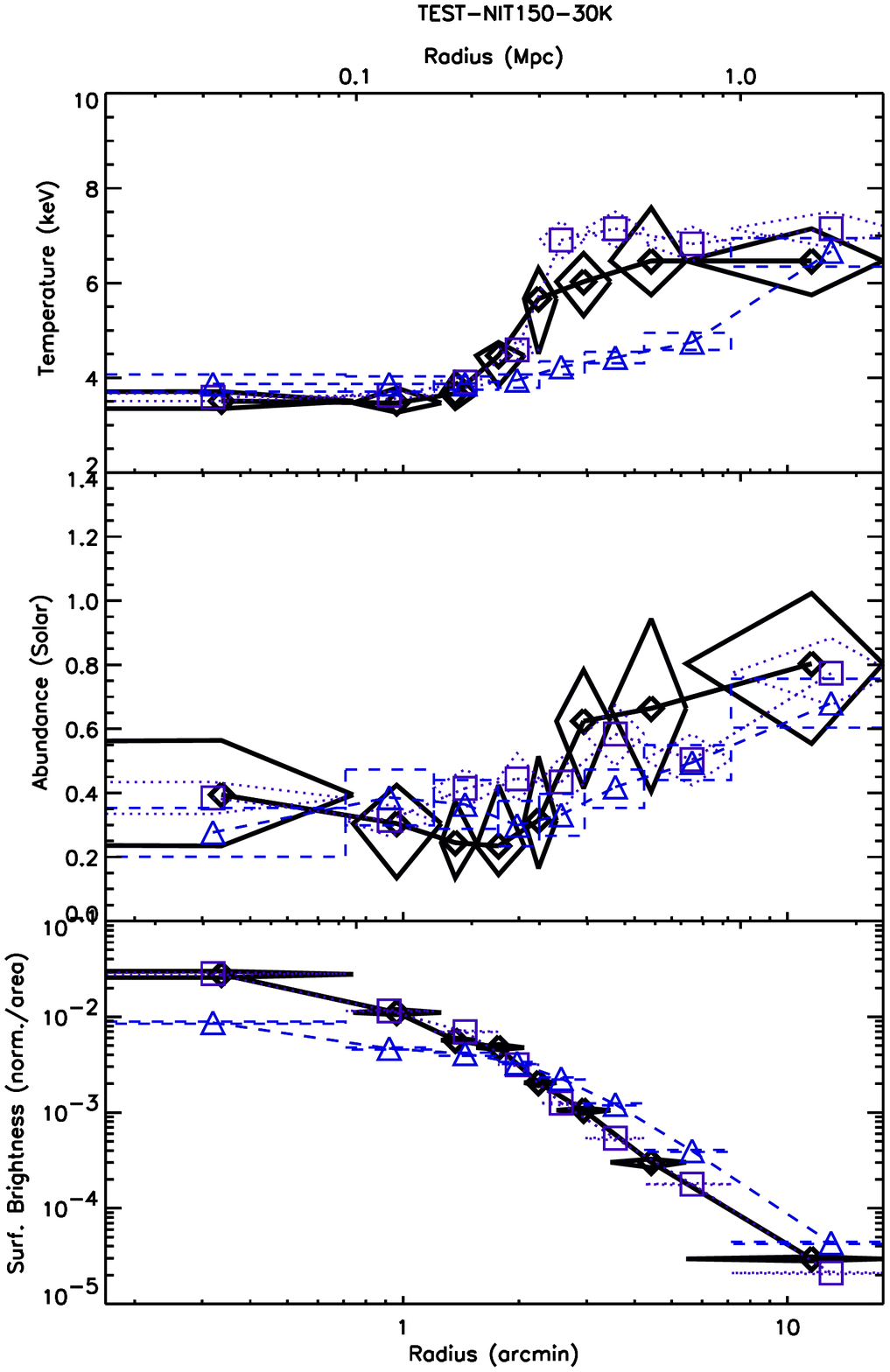,angle=0,width=0.49\textwidth,height=0.79\textheight}
		\centerline{(b)} } \parbox{0.99\textwidth}{
		\hfill\parbox{0.99\textwidth}{
		\captionfont\caption{\label{figure:testnit} This
		figure summarises the tests which utilize fewer (25)
		or greater (150) number of iterations in the M-L stage
		of the image deconvolution, compared to the standard
		number of 50.}}}
\end{figure*}
}

\def\testpsf{
\begin{figure}
	\parbox{0.49\textwidth}{
		\psfig{figure=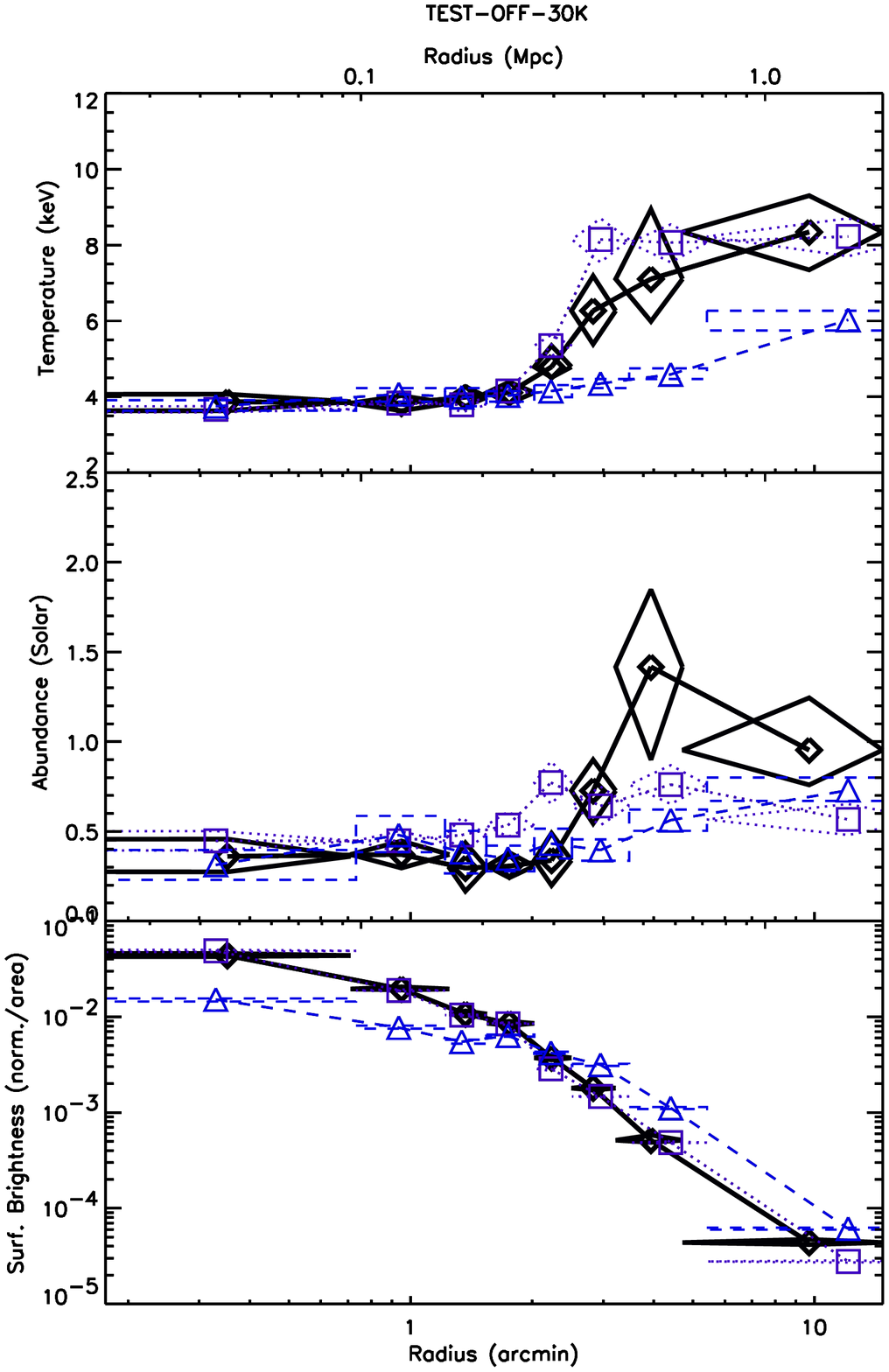,angle=0,width=0.49\textwidth,height=0.79\textheight} 
	}
	\parbox{0.49\textwidth}{
		\hfill\parbox{0.49\textwidth}{
		\captionfont\caption{\label{figure:testpsf} 
		This figure shows the limitation of using the
		fixed-spatial PSF in the \SID\/ procedure when the
		cluster is positioned at a large ($9\arcmin$) off-axis
		angle. 
	}}}
\end{figure}
}

\def\refsbr{

\begin{figure*}

  \parbox{0.99\textwidth}{
    \psfig{figure=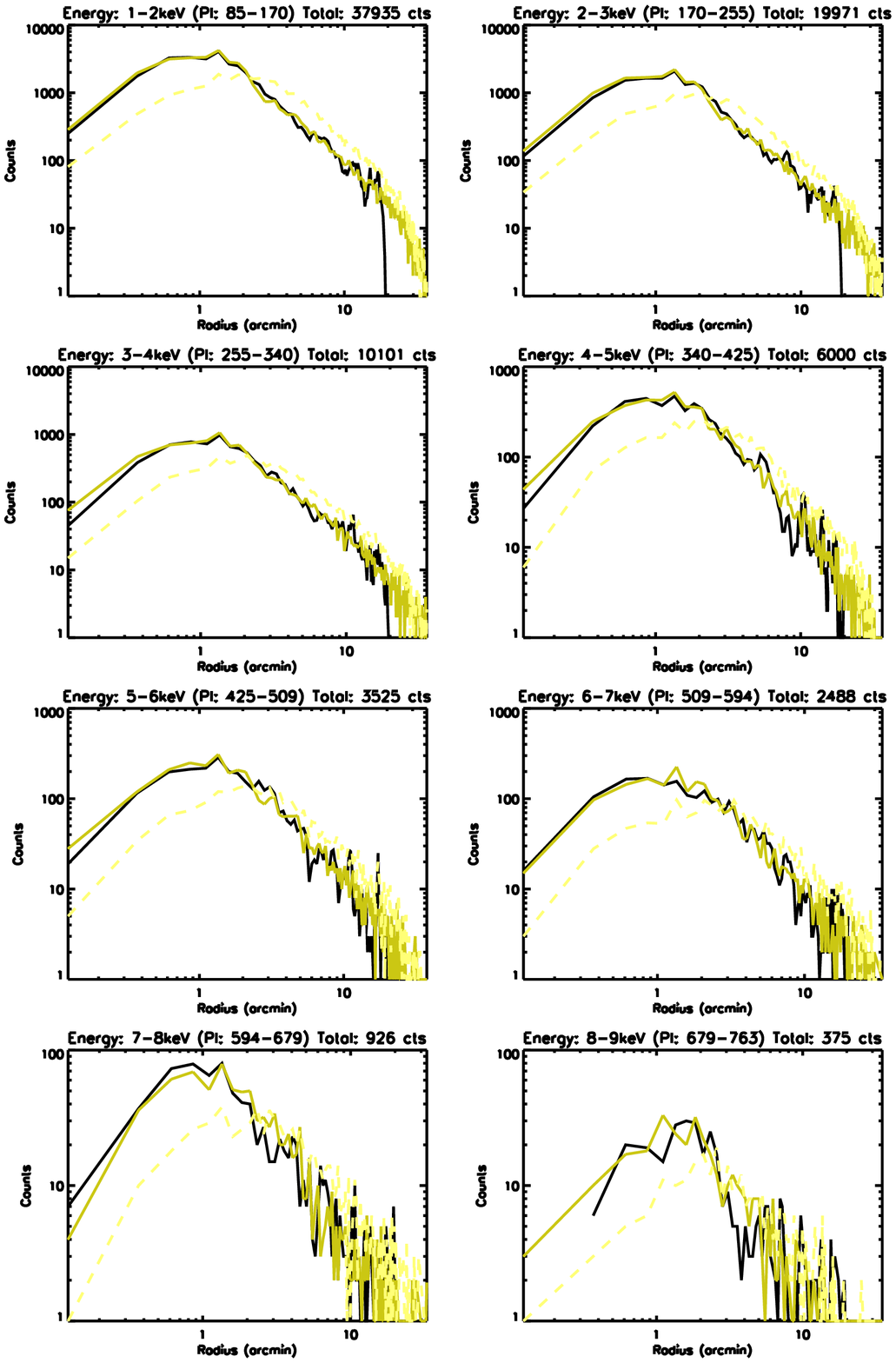,angle=0,width=0.99\textwidth,height=0.99\textwidth}
    } \parbox{0.99\textwidth}{ \caption{\label{figure:refsbr}} This
    figure presents a comparison of the original (medium-dark, solid
    line), deconvolved (dark, solid line) and convolved (light, dashed
    line) surface-brightness profiles for the (TEST-NOBGD-30K)
    simulations with no background contamination. This enables a
    direct comparison to be made of how well the \SID\/ procedure
    recovers the true surface-brightness profiles in each
    energyband. Although the \SID\/ procedure introduces extra noise,
    the true surface-brightness profiles are clearly recovered well
    (the deconvolved data are truncated outside $20\arcmin$), and are
    significantly better representations than the surface-brightness
    profiles from the convolved data, even in the higher-energy bands
    which have few counts. }

\end{figure*}

}

\def\flowchart{

\begin{figure*}

\parbox[]{0.9\textwidth}{\caption{\label{figure:flowchart} 
	Flowchart of the spectral-imaging deconvolution
	procedure. `SRC' refers to the `source plane' data (\ie\ the
	actual data before it is affected by the satellite
	instrumentation) which we are attempting to recover, and `IMG'
	refers to the `image plane' data (\ie\ the observed data after
	convolution\etc).}}

\begin{picture}(400,410)(-10,5)
\large

\put(40,405){\oval(50,10)}
	\put(0,400){\makebox(80,10){START}}
	\put(40,400){\vector(0,-1){10}}

\put(0,380){\framebox(80,10){Select energy band PSF (define PI range)}}
	\put(40,380){\vector(0,-1){10}}

\put(45,370){\makebox(80,10)[l]{(D) \em IMAGE DECONVOLUTION...}}

\put(0,360){\framebox(80,10){(D.1) Select events $\rightarrow$ in energy band IMG}}
	\put(40,360){\vector(0,-1){5}}

\put(0,345){\framebox(80,10){(D.2) Re-scale IMG \& PSF to same spatial scale}}
	\put(40,345){\vector(0,-1){5}}

\put(0,330){\framebox(80,10){(D.3) Deconvolve IMG with PSF $\rightarrow$ SRC}}
	\put(40,330){\vector(0,-1){10}}

\put(45,320){\makebox(80,10)[l]{(R) \em SPECTRAL REASSIGNMENT...}}

\put(0,310){\framebox(80,10){(R.1) Select individual PI channel in energy band}}
	\put(40,310){\vector(0,-1){5}}

\put(0,295){\framebox(80,10){(R.2) Create IMG image for this PI channel}}
	\put(40,295){\vector(0,-1){5}}

\put(0,280){\framebox(80,10){(R.3) Select random position from SRC image}}
	\put(40,280){\vector(0,-1){5}}

\put(0,265){\framebox(80,10){(R.4) Select random scattering by PSF}}
	\put(40,265){\vector(0,-1){5}}

\put(0,250){\line(4,1){40}}
	\put(0,250){\line(4,-1){40}}
	\put(80,250){\line(-4,1){40}}
	\put(80,250){\line(-4,-1){40}}
	\put(0,245){\makebox(80,10){(R.5) Corresponding event in IMG image?}}
	\put(40,240){\vector(0,-1){5}}
	\put(40,235){\makebox(10,5){YES}}
	\put(80,250){\vector(1,0){10}}
	\put(80,250){\makebox(10,5){NO}}

\put(0,225){\framebox(80,10){(R.6) Store event with IMG PI and SRC position}}
	\put(40,225){\vector(0,-1){5}}

\put(0,210){\line(4,1){40}}
	\put(0,210){\line(4,-1){40}} 
	\put(80,210){\line(-4,1){40}}
	\put(80,210){\line(-4,-1){40}}
	\put(0,205){\makebox(80,10){(R.7) ALL IMG events in this PI
	reassigned?}}

\put(80,210){\line(1,0){10}}
	\put(85,210){\makebox(0,5){NO}}
	\put(90,210){\line(0,1){75}}
	\put(90,285){\vector(-1,0){10}}

\put(40,200){\line(0,-1){5}}
	\put(40,195){\makebox(10,5){YES}}
	\put(40,195){\line(1,0){90}}
	\put(130,195){\vector(0,1){110}}

\put(90,315){\line(4,1){40}}
	\put(90,315){\line(4,-1){40}}
	\put(170,315){\line(-4,1){40}}
	\put(170,315){\line(-4,-1){40}}
	\put(90,310){\makebox(80,10){Done all PI channels in this band?}}
	\put(130,325){\makebox(10,5){YES}}

\put(90,315){\vector(-1,0){10}}
	\put(80,315){\makebox(10,5){NO}}
	\put(130,325){\vector(0,1){50}}

\put(90,385){\line(4,1){40}}
	\put(90,385){\line(4,-1){40}}
	\put(170,385){\line(-4,1){40}}
	\put(170,385){\line(-4,-1){40}}
	\put(90,380){\makebox(80,10){Done all energy bands?}}

\put(90,385){\vector(-1,0){10}}
	\put(130,395){\vector(0,1){5}}
	\put(80,385){\makebox(10,5){NO}}

\put(130,405){\oval(50,10)}
	\put(90,400){\makebox(80,10){STOP}}
	\put(130,395){\makebox(10,5){YES}}

\normalsize

\end{picture}

\end{figure*}

}

\def\variants{

\begin{table*}

\begin{tabular}{llccccccccccl}\hline

\multicolumn{1}{c}{Test Name} &
\multicolumn{1}{l}{Comments} &
\multicolumn{4}{c}{Systematic/Instrumental Flags} &
\multicolumn{1}{c}{} &
\multicolumn{5}{c}{Cluster Component} & 
\multicolumn{1}{c}{Fig.} \\ \cline{3-6} \cline{8-12} 

\multicolumn{1}{c}{} & 
\multicolumn{1}{l}{} & 
\multicolumn{1}{c}{Bgd.} & 
\multicolumn{1}{c}{Pos.} & 
\multicolumn{1}{c}{Exp.} & 
\multicolumn{1}{c}{$N_{\rm IT}$} & 
\multicolumn{1}{c}{} &
\multicolumn{1}{c}{$\Tx'$} & 
\multicolumn{1}{c}{$\dnH''$} & 
\multicolumn{1}{c}{$\Tx''$} & 
\multicolumn{1}{c}{$\Rcore''$} & 
\multicolumn{1}{c}{$R_{\rm max}''$} & 
\multicolumn{1}{c}{Ref.} \\ \hline 

{\em TEST-TAK-30K} &
	Takahashi \etal\ model &
	Yes   & 4.5   & 30    & 50    & ~     &
	9.87  & N/A   & N/A   & N/A   & N/A   &
	\ref{figure:testtak} \\
	
\\
{\em TEST-GCF-30K} &
	`Standard' model &
	Yes   & 4.5   & 30    & 50    & ~     &
	6.61  & 6.8   & 1.65  & 0.2   & 0.3   &
	\ref{figure:teststd} \\
	
\\
{\em TEST-MCF-30K} &
	Smaller cooling flow effect &
	$..$  & $..$  & $..$  & $..$  & ~     & 
	$..$  & $..$  & 3.31  & $..$  & $..$  &
	\ref{figure:testtx}(a) \\
	
{\em TEST-ISO-30K} &
	Isothermal &
	$..$  & $..$  & $..$  & $..$  & ~     &
	$..$  & N/A   & N/A   & N/A   & N/A   &
	\ref{figure:testtx}(b) \\

{\em TEST-DTX-30K} &
	Temperature decline &
	$..$  & $..$  & $..$  & $..$  & ~     &
	1.65  & 0     & 6.61  & 0.8   & 5     &
	\ref{figure:testtx}(c) \\

\\
{\em TEST-NOBGD-30K} &
	No background &
	No    & $..$  & $..$  & $..$  & ~     & 
	$..$  & $..$  & $..$  & $..$  & $..$  &
	\ref{figure:testbgd} \\

{\em TEST-GCF-15K} &
	Short exposure &
	$..$  & $..$  & 15    & $..$  & ~     & 
	$..$  & $..$  & $..$  & $..$  & $..$  &
	\ref{figure:testexp}(a) \\

{\em TEST-GCF-60K} &
	Long exposure &
	$..$  & $..$  & 60    & $..$  & ~     & 
	$..$  & $..$  & $..$  & $..$  & $..$  &
	\ref{figure:testexp}(c) \\

{\em TEST-NIT25-30K} &
	Less image deconv. iterations &
	$..$  & $..$  & $..$  & 25    & ~     & 
	$..$  & $..$  & $..$  & $..$  & $..$  &
	\ref{figure:testnit}(a) \\

{\em TEST-NIT150-30K} &
	More image deconv. iterations &
	$..$  & $..$  & $..$  & 150   & ~     & 
	$..$  & $..$  & $..$  & $..$  & $..$  &
	\ref{figure:testnit}(b) \\

{\em TEST-OFF-30K} &
	Off-axis cluster &
	$..$  & 9.0   & $..$  & $..$  & ~     & 
	$..$  & $..$  & $..$  & $..$  & $..$  &
	\ref{figure:testpsf}(b) \\

\hline
\end{tabular}

\parbox[]{0.9\textwidth}{

\caption{Variable Parameters For Generating Simulated Clusters
\label{table:variants}} In all but the single-phase temperature model
({\em TEST-ISO-30}), there are two gas phases, as indicated by the
primary and secondary superscripts. The model components which are
inapplicable (\ie\ the {\em TEST-ISO-30K\/} test) are indicated as
such by `N/A'.  Entries with `..' indicate the values default back to
the `standard model': {\em TEST-GCF-30K\/}.  Parameters which are
invariant between each test are: (i) the total column density (\ie\
Galactic; acting on both components where applicable):
$1.36\times10^{21}\psqcm$, (ii) redshift: 0.0881, (iii) baryon
fraction: 0.1, (iv) $\beta$-parameter: 0.67, (v) abundance:
$0.4Z_{\Solar}$, (vi) velocity dispersion: $900\kmps$ (both components
where applicable), (vii) core radius: $0.2\Mpc$ (primary component),
and (viii) maximum radial extent: $5\Mpc$ (primary component). Each
test is labelled to indicate the primary point in question. Those
tests which investigate different cluster temperature profiles are:
giant cooling flow: `{\em TEST-GCF-30K}'; moderate cooling flow: `{\em
TEST-MCF-30K}'; isothermal cluster: `{\em TEST-ISO-30K}'; radially
decreasing temperature: `{\em TEST-DTX-30K}'. Those tests
investigating instrumental/systematic effects are: inclusion of cosmic
background `{\em TEST-NOBGD-30K}'; off-axis position `{\em
TEST-OFF-30K}'; number of image-deconvolution iterations: `{\em
TEST-NIT25-30K}' and `{\em TEST-NIT150-30K}'; exposure time (in
kilo-seconds): `{\em TEST-GCF-15K}' and `{\em TEST-GCF-60K}'.  The
abbreviations for the the systematic/instrument flags are as follows:
`Bgd.': background included; `Pos.': off-axis position of the centre
of the simulated cluster; `Exp.': exposure duration; and `$N_{\rm
IT}$': number of iterations used in the M-L image deconvolution
stage. The abbreviations for the cluster properties are: gas
temperature `$\Tx$'; excess absorption (only on second component):
`$\dnH$'; core-radius: `$\Rcore$'; and truncation maximum radius:
`$R_{\rm max}$'. Finally `Fig.' indicates the figure reference for
each test, while each test is discussed in
Section~\ref{section:results}.

}

\end{table*}

}

\def\<{\thinspace}

\def\EINSTEIN{{\em Einstein}}

\def\EXOSAT{{\em EXOSAT}}

\def\ROSAT{{\em ROSAT}}

\def\ASCA{{\em ASCA}}

\def\GINGA{{\em Ginga}}

\def\Mpc{{\rm\thinspace Mpc}}

\def\cm{{\rm\thinspace cm}}

\def\deg{^\circ}

\def\eg{{\it e.g.\ }}

\def\etal{{et al.}}
\def\etc{{\it\thinspace etc.}}

\def\ie{{\it i.e.}}

\def\keV{{\rm\thinspace keV}}

\def\kmps{\hbox{$\km\s^{-1}\,$}}
\def\km{{\rm\thinspace km}}

\def\psqcm{\hbox{$\cm^{-2}\,$}}

\def\s{{\rm\thinspace s}}
\def\sec{{\rm\thinspace s}}
\def\ksec{{\rm\thinspace k\sec}}

\mathchardef\twiddle="2218

\def\arcmin{{\rm\thinspace arcmin}}

\def\Tx{\hbox{$T_{\rm X}\,$}}

\def\Rcore{\hbox{$R_{\rm core}\,$}}

\def\nH{\hbox{$N_{\rm H}\,$}}

\def\dnH{\hbox{$\Delta \nH\,$}}
\def\singlespace {\smallskipamount=3pt plus1pt minus1pt
                  \medskipamount=6pt plus2pt minus2pt
                  \bigskipamount=12pt plus4pt minus4pt
                  \normalbaselineskip=12pt plus0pt minus0pt
                  \normallineskip=1pt
                  \normallineskiplimit=0pt
                  \jot=3pt
                  {\def\smallskip {\vskip\smallskipamount}}
                  {\def\medskip   {\vskip\medskipamount}}
                  {\def\bigskip   {\vskip\bigskipamount}}
                  {\setbox\strutbox=\hbox{\vrule 
                    height8.5pt depth3.5pt width 0pt}}
                  \parskip 6.0pt
                  \normalbaselines}
\def\middlespace {\smallskipamount=4.5pt plus1.5pt minus1.5pt
                  \medskipamount=9pt plus3pt minus3pt
                  \bigskipamount=18pt plus6pt minus6pt
                  \normalbaselineskip=18pt plus0pt minus0pt
                  \normallineskip=1pt
                  \normallineskiplimit=0pt
                  \jot=4.5pt
                  {\def\smallskip {\vskip\smallskipamount}}
                  {\def\medskip   {\vskip\medskipamount}}
                  {\def\bigskip   {\vskip\bigskipamount}}
                  {\setbox\strutbox=\hbox{\vrule 
                    height12.75pt depth5.25pt width 0pt}}
                  \parskip 9.0pt
                  \normalbaselines}
\def\doublespace {\smallskipamount=6pt plus2pt minus2pt
                  \medskipamount=12pt plus4pt minus4pt
                  \bigskipamount=24pt plus8pt minus8pt
                  \normalbaselineskip=24pt plus0pt minus0pt
                  \normallineskip=2pt
                  \normallineskiplimit=0pt
                  \jot=6pt
                  {\def\smallskip {\vskip\smallskipamount}}
                  {\def\medskip   {\vskip\medskipamount}}
                  {\def\bigskip   {\vskip\bigskipamount}}
                  {\setbox\strutbox=\hbox{\vrule 
                    height17.0pt depth7.0pt width 0pt}}
                  \parskip 12.0pt
                  \normalbaselines}

\def\defaultspace{\singlespace}
\defaultspace

\def\captionfont{\normalsize}

\title[Spectral-Imaging Deconvolution Method] {Deconvolution of ASCA
X-ray data: I. Spectral-imaging method}

\author[D.A.~White and D.~Buote]{
\parbox[]{6.5in} {
	\large
	D.A.~White$^{1}$ \& D.A.~Buote$^{1,2,3}$ \\ 
	\footnotesize 
	$^{1}$Institute of Astronomy, Madingley Road, Cambridge CB3~OHA.
		(E-mail: daw@ast.cam.ac.uk)\\
	$^{2}$UCO/Lick, University of California at Santa Cruz, 
		Santa Cruz, CA95064, U.S.A.\\
	$^{3}$AXAF Fellow\\
	}
}

\date{Received ***; in original form ***}

\defaultspace
\def\Tx{\hbox{$T_{\rm X}$}}

\def\bfit{\hbox{$\beta_{\rm fit}$}}
\def\bspec{\hbox{$\beta_{\rm spec}$}}

\def\Solar{\hbox{$\odot$}}

\def\MFSV{{\em {MFSV}}}
\def\SID{{\em {SID}}}

\def\nsample{106}

\def\nfrac{0.1}
\def\itmax{50}

\begin{document}

\maketitle 

%-------------------------

\begin{abstract}   

In this paper we describe a self-contained method for performing the
spectral-imaging deconvolution of X-ray data on clusters of galaxies
observed by the \ASCA\/ satellite. Spatially-resolved spectral studies
of data from this satellite require such a correction because its
optics redistribute photons over regions which are of comparable size
to the angular scales of interest in clusters. This scattering is a
function not only of spatial position but also energy. To perform a
correction for these effects we employ Maximum-Likelihood
deconvolution of the image (within energy bands of $1\keV$) to
determine the spatial redistribution, followed by a Monte-Carlo energy
reassignment of photon energies with position to determine the
spectral redistribution. We present tests on simulated cluster data,
convolved with the various instrumental characteristics and the X-ray
background, which show that our methodology can successfully recover a
variety of intrinsic temperature profiles in typical observational
circumstances. In Paper-II we apply our spectral-imaging deconvolution
procedure to a large sample of galaxy clusters to determine
temperature profiles, some of which will be used in subsequent mass
determinations, presented in Paper-III.

%Utilising our simulations we also show that we can produce the same
%systematic temperature declines as observed by
%\citeANP{Markevitch:ASCA_Tx_similarity} if we make an assumption
%similar to that which they employ in their methodology.

\end{abstract} 


%-------------------------

\section{Introduction}

One of the most important goals in the X-ray study of clusters of
galaxies is the determination of their mass properties. The radial
distribution of the total gravitating mass in a cluster can be
constrained by accurate spatially-resolved observations of the gas
temperature and density, combined with the simple assumption of
hydrostatic equilibrium. While the accurate determination of density
profiles is well within the capabilities of previous instrumentation
(\eg \ROSAT), accurate radial temperature profile constraints requires
good spectral {\em and\/} spatial resolution. 

The \ASCA\/ satellite has the spectral resolution required, but the
spatial resolution provided by its nested foil mirrors is inadequate
for the analysis of extended sources, such as galaxy clusters. The
on-axis half-power diameter of the point-spread function (PSF), of
approximately $3\arcmin$, increases and becomes increasingly
asymmetric with off-axis position. However, the essential problem with
the PSF is that it varies with energy, such that scattering is more
severe for harder photons. Consequently, if no correction is applied
to an \ASCA\/ dataset then the determination of the spatially-resolved
temperature characteristics of a cluster will be incorrect. For a
truly isothermal cluster, the PSF will produce a temperature profile
which appears to rise with increasing radius from the cluster core
\cite{Takahashi:ASCA_PSF}.

% -- see the ``\ASCA\/ ABC Guide''\nocite{ASCA:ABC})

Most of the results presented currently in the literature, that
attempt to correct for the spatial and energy-dependent nature of the
\ASCA\/ PSF, use a method created by \citeN{Markevitch:ASCA_A2163_hydeqm}. 
Their procedure relies on having an accurate prescription for the
emissivity profile of the cluster -- usually a $\beta$-model fit
obtained from a \ROSAT\/ surface-brightness profile. Simulated cluster
photons are generated according to an initial guess for the cluster
spectrum, and spatially distributed according to the emissivity
profile. The photons are then convolved with the spatially-variable
PSF in different energy bands, and the model cluster compared with the
observational data. The source spectrum is then varied at different
positions until the model and observational data are statistically
consistent. 

The advantage of their method is that the spatially-variable nature of
the PSF is easily incorporated, while the disadvantages are that the
results may be compromised by the applicability of the spectral model
and the accuracy of the emissivity profile. The latter issue may be
important because the \ROSAT\/ energy band ($0.2-2\keV$) is
significantly different from that of \ASCA\/ [$0.5-10\keV$ -- although
\citeN{Markevitch:ASCA_Tx_similarity} use $2-10\keV$]. Therefore,
there may be significant mismatch between the emissivity profile
described by \ROSAT\/ and that required for the
\ASCA\ data analysis, if the cluster exhibits spatial variations in
temperature or other properties which affect the spectrum, \ie\/ a
cooling flow cluster will have many temperature components and a
probable variation in the column density across the cluster (although
this has to be large to affect the spectrum above $2\keV$).

Using the above method, \shortciteN{Markevitch:ASCA_Tx_similarity}
(hereafter \MFSV) and others [\citeN{Markevitch:ASCA_A2163_hydeqm};
\citeN{Sarazin:ASCA_A2029};
\citeN{Donnelly:ASCA_A1367};
\citeN{Henricksen:ASCA_A754};
\citeN{Markevitch:ASCA_TriAus};
\citeN{Markevitch:ASCA_three_clusters};
\citeN{Markevitch:ASCA_four_hot_clusters}]
have presented temperature profiles for many clusters of galaxies. In
a particular \MFSV\/ analysed a sample of 30 objects and found that
most of these clusters have temperature profiles that decline with
radius, which they parameterised with an average polytropic index of
$\gamma=1.24^{+0.20}_{-0.12}$. 

This index is close to, and consistent with at the 2-$\sigma$ boundary
of their uncertainties, the convective instability limit of
$\gamma\ge5/3=1.67$. While convective instability may be expected in
clusters which have been disturbed, this should not be the case for
relaxed clusters. Indeed, because cooling flows should be disrupted in
significant merger events they can be taken as an indicator of cluster
relaxation (\citeNP{Buote:morphology_ii} and
\citeNP{Buote:Omega_substructure}). However, as \MFSV\/ themselves find
that approximately 60 percent of their clusters contain a cooling
flow, there is a inconsistency between the apparent proportion of
relaxed clusters in their sample and the possibility that the sample
is close to convective instability -- as implied by their steep
temperature gradients.

Additionally, the steep temperature gradients exacerbate the baryon
problem in clusters (\citeNP{White:baryon_catastrophe};
\citeNP{Briel:Coma_RASS}; \citeNP{White:baryon}) -- \ie\/ the apparent
discrepancy between the large relative fraction of the total mass in
baryons (\ie\/ essentially that seen as X-ray emitting gas), and the
fraction expected from the production of baryons during primordial
nucleosynthesis in a flat ($\Omega_0=1$) Universe. Declining
temperature gradients imply smaller total cluster masses than would be
inferred from isothermal temperature profiles, (if the electron and
ion temperatures are in equipartition, \citeNP{Ettori:Coulomb}),
thereby leading to even larger baryon fraction determinations.

Although \citeANP{Markevitch:ASCA_Tx_similarity} themselves show that
their average temperatures determinations agree well with those
obtained from previous broad-beam satellites such as \EINSTEIN,
\EXOSAT\/ and \GINGA, the crucial question is whether the
shape of their temperature profiles are correct. For this their
supporting evidence relies mainly on the consistency of the
temperature profiles from the different \ASCA\/ detectors.
\citeN{Markevitch:ASCA_three_clusters} have performed a comparison of
\ASCA\/ results with \ROSAT\/ temperature profiles, and found reasonable
agreement, but the \ROSAT\/ temperature determinations have rather
large uncertainties.

There are a growing number of results published which use different
methods. Many of these results (\ie\/
\citeNP{Ikebe:ASCA_Hydra-A}: A780;
\citeNP{Fujita:ASCA_A399andA401}: A399 and A401;
\citeNP{Ezawa:ASCA_AWM7}: AWM7;
\citeNP{Ohashi:ASCA_four_clusters}: 3A0336+098, MKW3s, A1795 and
PKS2354$-$35) have been compared with those from Markevitch \etal\/ by
\citeN{Irwin:ROSAT_Tx_profiles}. They highlighted the fact that many of these
other authors determine that these clusters have isothermal
temperature profiles, even in cases where Markevitch \etal\ find a
clear decline. \shortciteANP{Irwin:ROSAT_Tx_profiles} also presented
their own analysis of \ROSAT\/ colour profiles which showed that their
clusters were generally consistent with isothermality. Most recently,
\cite{Molendi:SAX_A2319} have presented Beppo-SAX results which show
that A2319 is isothermal, although these authors are reluctant to
claim any significant discrepancy with the Markevitch \etal\ results
on the basis of this one observation.

Given the importance of generic temperature declines in clusters, and
their contradiction with other results, it is essential that the
\ASCA\/ data are analysed by independent means to check the \MFSV\/
results. Also, \ASCA\/ data currently still provide the best
opportunity for the most accurate temperature profiles determinations
for a large sample of clusters. Thus, we we have created our own
`spectral-imaging' deconvolution (\SID) procedure. Our method is
self-contained: it requires only \ASCA\/ observational data on the
cluster, background, and the energy-dependent PSF. It does {\em not\/}
require \ROSAT\/ constraints on the emissivity profile. Also, no
assumption about the various components contributing to the source
spectrum is required in order to perform the deconvolution, and our
method is essentially non-parametric. The main assumption that we have
made is that a spatially {\em invariant\/} PSF is sufficient for our
purposes. This has been done for computational speed and for
convenience (we use a third party image deconvolution routine -- see
Section~\ref{section:mlimage}), although the consequences of this
assumption are tested for, and are shown to be acceptable.

Such tests are central to our methodology, in order that we may be
confident in the validity of our results. Thus, we have tested our
procedure on simulated clusters by subjecting them to the various
instrumental and background effects that will occur in the practical
application of the method to real GIS data. For example, we have
investigated the effect of the position-dependent nature of the PSF,
looked at the degradation in performance with decreasing
signal-to-noise, and tested the procedures ability to recover a
variety of intrinsic cluster temperature profiles.

In summary, this paper presents a spectral-imaging deconvolution
method for practical use with \ASCA\/ data on clusters of galaxies. In
Paper-II (\citeNP{White:deconv_ii}), we apply this method to a large
sample of \ASCA\/ GIS cluster observations, and compare our results
with the results from \MFSV.

%-------------------------

\section{Method}

Our procedure for the spectral-image deconvolution (\SID) of \ASCA\/
data can be broadly divided into two sections: image deconvolution
followed by spectral reassignment, as summarised by a flow-diagram of
the method shown in Fig.\ref{figure:flowchart}.

\subsection{Image Deconvolution}\label{section:mlimage}

The image deconvolution is performed in the Interactive Data Language
(IDL) environment using a maximum-likelihood (M-L) procedure,
distributed with the ASTRON\footnote{\tt
http://idlastro.gsfc.nasa.gov/homepage.html} library of contributed
routines by W.~Landsman. Although this procedure accounts for the
effect of Poissionian noise in the image, it has two significant
drawbacks: it assumes, (i) that the PSF is spatially invariant, and
(ii) that the data are monochromatic.

The first issue of the spatial variance of the PSF is neglected, as we
will assume that the PSF appropriate to the position of the cluster in
the detector image can be used for the whole image deconvolution. We
argue that this is reasonable, on the basis that most of the photons
we are interested in arise from the core of the cluster. However, the
impact of this assumption is assessed in the test detailed in
Section~\ref{section:testpsf}.

The monochromatic issue is dealt with by dividing the data according
energy, and executing the deconvolution on the image for each energy
band. This can not be done for the data in each individual PI (pulse
invariant) channel, as the M-L deconvolution procedure requires more
counts to work with than will be found within one PI channel, and the
PSFs which are available in the \ASCA\/ calibration database (CALDB)
at HEASARC (High-Energy Astrophysics Archive, Goddard Space Flight
Center) are only supplied in $1\keV$ bandwidths from 1 to $10\keV$.

Our methodology is to take the X-ray events from each cluster
observation and divide them into broad spectral energy bands,
corresponding to the energy ranges of the PSF images. We then
construct images from the data events in each of these energy bands
(Fig.\ref{figure:flowchart} -- steps D.1 and D.2) and then pass them,
along with the appropriate point-spread function, to the image
deconvolution procedure (step D.3). After \itmax\ iterations the M-L
procedure returns the deconvolved image for that energy band (step
D.4). (The effect of varying the number of iterations is investigated
in Section~\ref{section:testnit}.)

These deconvolved images are then supplied to the spectral
resassignment phase to determine the PI energy of events which will
comprise the deconvolved dataset.

\subsection{Spectral Re-assignment}\label{section:mlspectral}

From the M-L procedure we know the spatial probability distribution of
photons in the deconvolved plane. We also know that the total number
of counts are conserved between the convolved and deconvolved planes.
Therefore, we can create a random set of positions and energies from
the deconvolved image and the observed PI distribution. However, we do
not know which event has which PI energy.

Instead of attempting to recover the individual PI information for
each event we could have performed a colour analysis of the
deconvolved surface-brightness profiles in each $1\keV$ energy band
(approximately 85 PI channels). However, this would fail to utilise
the superior spectral capabilities of \ASCA, and would lead to a
degeneracy in temperature and abundance determinations. Thus, we have
endeavoured to maintain the spectral resolution of \ASCA\/ by finding
a way of assigning energies to the events list created from the
deconvolved images.

The obvious procedure would be to randomly assign an energy for each
event using the overall spectral distribution of events within the
$1\keV$ energy band, however this is not correct if there is spatial
variation of the PI events over this energy band, as the following
example illustrates. Consider two spatially distinct point sources
within a single $1\keV$ energy band interval: one emitting $1.2\keV$
photons and the other $1.8\keV$ events. A random assignment of
energies in this energy band, which does not take into account any
spatial information, will result in each point source having half of
the $1.2\keV$ photons and half of the $1.8\keV$ photons. If the same
number of photons were detected from each source, they would both end
up with an incorrect mean energy of $1.5\keV$.

To make the correct reassignment we need to choose a `prior' spatial
distribution for the events within each $1\keV$ energy band. For this
we use the M-L deconvolved image (while \MFSV\ use the \ROSAT\/ data
to obtain a higher spatial-resolution emissivity profile). We can then
effectively `ray-trace' events from this image through the telescope
optics and determine the most likely association between events in the
deconvolved and convolved planes. As we know the energies of each
convolved event we then know the energy of the deconvolved event.

In our ray-tracing method we assume the PSF is invariant within each
energy band. (We discuss the systematic bias that this can introduce
into our results in Section~\ref{section:syspsf}.) For each PI channel
we randomly generate the appropriate number of {\em source}-plane
events from the deconvolved image (steps R.1 to R.3). We then use the
PSF to scatter these events into the {\em image}-plane (step R.4). If
any of these {\em image}-plane events correspond to the positions of
an observed photon (for the PI energy in question), then we consider
this `mapping' to be successful (R.5). We do this until all the
photons in the PI channel in question have been successfully mapped.

Having obtained the position and energy of a {\em source}-plane event
we can store the information (R.6) and eliminate the appropriate
photon from the observed event list. We continue the ray-tracing until
all the observed events are accounted for (R.7). As the number of
events to be reassigned declines, the probability of finding a
successful mapping reduces. For efficiency we mask out parts of the
PSF which can not provide a successful link between a {\em source\/}-
and {\em image}-plane position. Each randomisation is thereby
guaranteed a successful mapping and the computational efficiency is
improved significantly.
	
%Although this spectral reassignment can be applied to two-dimensional
%data, we have reduced the problem to one-dimension to aid
%computational speed and coding simplicity. More specifically, the
%image deconvolution is performed in two-dimensions but the spectral
%reassignment in executed in one-dimension. Thus, azimuthally averaged
%{\em counts\/} profiles, centred on the peak of the cluster emission,
%are extracted for the deconvolved and observed (\ie\/ convolved) data;
%the PSF is also converted to a counts profile. As conventional
%spectral analysis requires the event details in detector coordinates,
%the one-dimensional result is converted back to a two-dimensional
%image. The events listing can then be analysed in the same manner as
%conventional data, however it should be remembered that the data now
%implicitly incorporate the assumption of spherical symmetry.

\subsection{Practical Considerations}\label{section:mcarlo}

	\flowchart
	\psffig
	\refsbr

We have implemented our procedure in the Interactive Data Language
(IDL) environment. The deconvolved events listings are written to
standard (FITS) format files, which are then analysed using
conventional X-ray data processing procedures (FTOOLS 4.2 and XSPEC
10.00 -- see below). We currently only analyse GIS data as it has a
larger field-of-view (FOV)\footnote{The usable area of the GIS is
circular and $\sim40\arcmin$ in diameter, while the SIS is square and
$\sim18\arcmin$ on a side.} than the SIS -- which allows us to
determine cluster properties to larger radii. (The larger FOV also
conserves more of the photons scattered to larger off-axis angles by
the PSF. The SIS also has gaps between the CCDs, and the individual
CCDs have differing instrumental gains.)

With the methodology we have defined above, there are various
assumptions and complications which have to be assessed before we can
be confident in applying it to real data. Firstly, we have noted, the
M-L image deconvolution procedure only accepts a spatially invariant
point-spread function, while the X-ray telescope (XRT) of \ASCA\/
exhibits a significant variation with position around the field of
view (\eg\ see Fig.~\ref{figure:psffig}). In the spectral reassignment
we also assume that the PSF does not change significantly across each
$1\keV$ energy band. Secondly, observational data also include
contaminating events from cosmic X-ray sources, and from the X-ray
detector itself. Finally, systematic effects introduced by the
procedure have to be minimised.

Dealing with the last of these points first, our experience has shown
that a single realisation of the deconvolved data is subject to
systematic effects. In particular, we noticed that the abundance
profiles show abrupt variations which were clearly due to systematics
in the procedure (the M-L image deconvolution). However, these can be
reduced by repeating the \SID\ procedure many times. Due to time
constraints, we found that 10 repetitions was the maximum number we
could deal with practically. To perform the Monte-Carlo procedure we
randomise each input (convolved) dataset, and then process each
deconvolved events list independently in separate spectral
analyses. The spectral results are then combined to give the average
profiles presented in the simulation results plots (this
radial-profile averaging procedure is described in Paper-II). As the
error bars in these plots are larger than the variations from
bin-to-bin (and larger than the errors on either the original or
convolved data) it is clear that the errors in the deconvolved
profiles are dominated by systematic errors rather than statistical
errors. However, the abundance profiles, in particular, are now much
better behaved than found in any single realisation.

For the other two points listed above, we believe the only effective
way to investigate the validity of our procedure, especially in
circumstances typical of an \ASCA\/ observation, is to create
simulated cluster datasets and then apply the various observational
and systematic effects to see how well the original physical
properties of the intracluster gas are recovered by the deconvolution
procedure. In the description below, we indicate how we create these
various simulations and how we assess the impact of these various
contaminants in practical observational circumstances.

%-------------------------

\section{Analysis}\label{section:analysis}

\subsection{Generating Cluster Models}
\label{section:generation}

We generate the basic simulated cluster dataset from a radial density
profile, parameterised by a standard $\beta$-model, and an isothermal
temperature profile. The expected flux from each component (a second
component is added to mimic a cooling flow cluster) of the
intracluster medium (ICM) is determined by projecting the volume
emissivity along the line-of-sight and convolving this with the
spectral response of the detector. The distribution of counts in each
PI detector channel is then obtained by randomly sampling the
intrinsic cluster spectrum. This is repeated until the total observed
flux is obtained for each thermal component. Similarly, the detected
spatial position of each event is determined by random generation from
the projected emissivity, followed by convolution with the spatial
PSF. Note, the effect of vignetting is implicitly incorporated by our
use of a spatially variable PSF to create the simulated datasets, as
the normalisation of the PSF describes the efficiency of the detector
at that position.

The cosmic background (from observations held at the \ASCA\/ CALDB at
HEASARC) can also be included by adding in the expected number of
background events from an observation of a blank area of real sky, and
scaling for the relative exposure time of the simulated
dataset. Because the background data are taken from actual
observations of blank-sky fields these data are already convolved with
the real spatially-variable PSF and include all the various sources of
noise, such as the GIS calibration source, \eg\/ see
Fig.~\ref{figure:imgfig}(b).

	\imgfig

The convolved simulated dataset, incorporating optional instrumental
and background effects, is then supplied to our \SID\/ procedure,
along with the PSFs. The PSFs in each $1\keV$ energy band are selected
to most closely correspond to the position of the center of the
cluster in the detector image. In the following section, we compare
the results from the various tests which are designed to assess the
impact of the assumptions implicit in our method, and the various
observational contaminants such as the X-ray background.

We judge the success of the \SID\/ procedure by its ability to recover
the properties of the original simulated cluster, and to do this we
perform a spatially-resolved spectral analysis on the original,
convolved, and deconvolved events and then compare the results
graphically. Although the correct recovery of the original temperature
profiles by the \SID\/ procedure implicitly requires that the original
surface-brightness profiles are correctly recovered by the M-L
deconvolution in each $1\keV$ energyband, this is explicitly shown in
Fig.~\ref{figure:refsbr}.

\subsection{Spectral Analysis}
\label{section:spectra}

The spectral analysis procedure which we employ is fairly standard. We
start by defining annular regions of interest, centred on the peak of
the X-ray emission, and then fit a spectral model to each of these
regions and compare the results. The maximum radius allowed for any
cluster extraction is $20\arcmin$ from the centre of the field of view
(although the region of most practical interest will generally be
within the inner $10\arcmin$). This eliminates the gain uncertainties
at the edge of the GIS detector and contamination from the calibration
source. However, if the background has been incorporated the maximum
radius is usually reduced to the radius where the
background-subtracted surface-brightness profile remains positive. The
background spectrum is extracted from the blank sky observations using
the same spatial regions defined by the cluster data. In this way any
gross position-dependent detector variations, such as instrument gain,
should cancel out.

After defining each annular region, such that each annulus contains a
certain fraction of the total (if applicable -- background subtracted)
counts (\nfrac\/\footnote{The nominal fraction of the total number of
background-subtracted counts per annulus is 0.1, the soft-limit on the
minimum number of counts in an annulus is 1,000 and the hard-limit is
500 counts.} -- which gives a theoretical maximum of 10 radial bins;
see Paper-II for details on the methodology), we extract events using
FTOOLS~4.2. This is done for both the source and background
spectra. These spectra, together with the ancillary response matrix
(\ie\/ ARF file; determined using ASCAARF), are passed to the
XSPEC~10.00 spectral analysis package (\citeNP{Arnaud:XSPEC}), and
fitted with the chosen spectral model.  We note that, until this
point, no assumption has been made about the intrinsic cluster
spectrum, and any chosen model can be fitted to the deconvolved data
-- contrary to the procedure used by \MFSV\/ which requires a spectral
model to be defined as an integral part of the deconvolution.

In this analysis we fit a single temperature plasma emission
model\footnote{Our specific model is that of a single MEKAL thermal
component (\ie\/ \citeNP{Mewe:MEKALa}; \citeNP{MEWE:MEKALb}), absorbed
(\citeNP{Morrison:wabs}) by a foreground column density of foreground
material.} -- regardless of whether the simulated data were created
with two thermal components. This means that we are not using the
correct model for the data, but for the simulations this is not
important. The crucial fact is whether we recover the same results
from the deconvolved data as from the original data using the same
spectral model. Thus, in the following results we are looking for good
agreement between the spectral fits to the deconvolved and original
datasets (\ie\/ simulated data which have been convolved with the
spectral response, but not the spatial PSF).  (We do note, however,
that our lack of a cooling flow component, and particularly the excess
absorption, in the fitted model is largely responsible for the radial
increase in metallicities in the cooling flow simulations, which is
not apparent in the single thermal component models.)

	\testtak

Finally, we note that if background contamination is included in the
test it is also applied to the convolved data (and thereby the
deconvolved data), but not the original data.  Thus, no background
subtraction is performed in the spectral analysis of the original
dataset. This also means that (except in the test which has no
background added) the convolved and deconvolved results are subject to
extra noise which is not in the original dataset spectral fits.

\section{Cluster Simulation Results}\label{section:results}

\subsection{Takahashi \etal\ Model -- The Effect Of The PSF On ASCA Cluster Data}
\label{section:testtak}

In the Introduction we referred to a paper by
\cite{Takahashi:ASCA_PSF} which showed that the energy-dependent PSF
of ASCA can give rise to an apparently increasing temperature profile
in an isothermal cluster, if the data were not corrected for the
effect of the PSF. Before we discuss our various simulation results,
where we investigate different observational and physical scenarios,
we show that we can reproduce this systematic effect when using
similar parameters. (The magnitude of the effect depends on
parameters, such as apparent size of the cluster core.)

As the effect of the PSF becomes severer for hotter clusters, we chose
to approximate the hottest temperature profile modeled by
\citeN{Takahashi:ASCA_PSF}. We do this using a single $\beta$-model
with a velocity dispersion of $1,100\kmps$ (which with the
$\bspec=0.8$ which we use in our other models gives a central
temperature of approximately $9-10\keV$), an index for the profile of
$\bfit=0.6$ and a core radius of $\Rcore=0.13\Mpc$ (resulting in a
$1\arcmin$ core, similar to \citeANP{Takahashi:ASCA_PSF}). Further
details of the parameters used are given in Table~\ref{table:variants}
under the {\em TEST-GCF-30K\/} model.

In Fig.~\ref{figure:testtak} the effect of the PSF can be seen in the
convolved dataset's temperature profile (dashed line with triangle
symbols), where the temperature in the core is low and then it
gradually increases with radius. This trend is similar to the
\citeANP{Takahashi:ASCA_PSF} profile, which we present in the plot as
square symbols with dashed error-bar lines. The diagram also shows
that our \SID\ method recovers the true radial temperature,
metallicity and emissivity profiles from the convolved data.

In the following tests we create a suite of models to investigate the
effects of various observational and physical conditions on the
ability of the \SID\ procedure to recover the true intracluster gas
properties.

\subsection{Standard Test -- A Cluster With A Strong Core Temperature Decline}
\label{section:teststd}

	\variants % see incl.tex

	\teststd

With our basic prescription for creating a simulated cluster dataset
(see Section~\ref{section:generation}), we choose our standard model
to represent a `Giant Cooling Flow' ({\em TEST-GCF-30K}). The
variation in the average temperature in this type of cluster is
expected to be significant, and will therefore provide a severe test
for the \SID\/ procedure. To produce a core temperature decline we add
a second cooler thermal component to the core region. Although this is
not a physically realistic representation of a cooling flow it mimics
a core temperature drop (albeit rather sharply) which is seen in
typical cooling flow clusters (\eg\/ Centaurus --
\citeNP{Allen:Centaurus_PSPC}). (Remember, we are not necessarily
interested in the physical correctness of our model but whether the
\SID\/ recovers the intrinsic energy and spatial distribution of the
simulated cluster -- whatever that might be.)

The parameters have been chosen to approximate the physical
characteristics of Abell 478. This is a cluster of moderate
temperature, and distance, but is fairly luminous due to its large
cooling flow. It should be remembered that some of the conclusions
drawn from the following tests will be dependent on the
characteristics of the cluster in question, \eg\/ whether the cluster
has a cooling flow, or not, and how bright the cluster is, etcetera.
Table~\ref{table:variants} shows the parameters we have use in each
test. The primary component has a temperature of $6.6\keV$; the cooler
component is at $1.65\keV$, but its radial extent is truncated at
$1.5\times\Rcore=0.3\Mpc$ to limit it to the centre of the
cluster. The cooler component spectrum is also modified by an
additional column of absorbing material of five times the assumed
Galactic value (which is set at $1.36\times10^{21}\psqcm$) to model
excess absorption (\ie\/ \citeNP{White:SSS_abs} and
\citeNP{Johnstone:CF_A478}). For this standard model, the
observational characteristics are an exposure of $30\ksec$, with the
centre of the cluster placed $4.5\arcmin$ off axis (fairly standard
for an \ASCA\/ GIS observation), and the data are convolved using a a
spatially varying PSF. (The PSF is interpolated from the grid of 11
different positions for the PSF in each energy band, supplied in the
CALDB.)

%In the M-L image deconvolution we use \itmax\ iterations.
%The count rate for the cluster in this standard model, and the
%subsequent models, are given in Table~\ref{table:simdat} -- under the
%model labelled {\em TEST-GCF-30K}.

%The results obtained from a spectral analysis of the whole cluster in
%the original, convolved, and deconvolved datasets are summarised in
%Table~\ref{table:simresall}.
%The broad-band spectral results show that there is no real difference
%between the overall characteristics as determined from the original,
%convolved or deconvolved datasets. 

The effect of the spectral-image deconvolution on the convolved data
becomes apparent when we look at the radial variations in the spectral
fit parameters in Fig.~\ref{figure:teststd}. This figure shows that
the variations of the original ICM properties are recovered by the
\SID, and are a much better representation of the original variations
than those obtained from the convolved data. (If a hydrostatic mass
determination were made from the convolved data on such a cluster,
then careful consideration of the complications would be required, if
meaningful results were to be obtained.) We note, however, that the
abruptness of the temperature profile decline in the original data,
which is rather artificial due to the construct of our model, is
somewhat smoothed over in the deconvolved temperature profile.

\subsection{A Variety Of Different Temperature Profiles}
\label{section:testtx}

In the next three tests we have attempted to span a wide range of
possible temperature profiles. With the `Medium Cooling Flow' model,
{\em TEST-MCF-30K}, we investigate whether a less dramatic temperature
decline can be recovered.  In this case, the simulated data were
created by adding a secondary component which is slightly hotter,
\ie\/ $3.3\keV$, than in the previous model. Figure~\ref{figure:testtx}(a)
shows that the small differences between the original and convolved
temperature profiles are still resolved with the deconvolved data.

In the {\em TEST-ISO-30K\/} model, Fig.~\ref{figure:testtx}(b), we
eliminate the second cooler component altogether, and produce an
isothermal temperature profile. This test shows that the deconvolution
procedure essentially has no systematic effects which severely distort
the recovery of a flat temperature profile and, for example, results
in a temperature decline.

	\testtx
	\testtxcontd

In the last temperature profile variation, {\em TEST-DTX-30K} -- shown
in Fig.~\ref{figure:testtx}(c), we have created a temperature profile
which decreases radially outwards, to approximate the general trend
for decreasing temperature profiles found by \MFSV. This model was
created by making the hotter component more extended (\ie\/ the
core-radius is larger), than the cool component (note, the truncation
is set at $5\Mpc$ for both components and there is no excess
absorption on the cooler emission component). Although the difference
between the original and convolved temperature determinations in any
particular radial bin is small, the general slope of the original
profile is more accurately reproduced by the deconvolved data.

Despite the lower count rates from the {\em TEST--MCF--30K}, {\em
TEST--ISO--30K} models, as they incorporate less flux from the
secondary temperature component, these results show that the success
of the deconvolution does not appear to be dependent on the intrinsic
temperature variations, and that there are no severe systematic biases
in the deconvolved profiles.

\subsection{Statistical Effects Due To Background And Exposure}
\label{section:testbgd}

We now turn our investigation to the impact of the various
instrumental and systematic effects which occur in \ASCA\/
observations of clusters: the effect of background event
contamination, and the impact of signal-to-noise variations.

In {\em TEST-NOBGD-30K} we see the intrinsic ability of the \SID\/
procedure in the absence of any background contamination. The results
are clearly more accurate, but are still qualitatively similar to the
results from the standard simulation data which includes the
background. This shows that performance of the \SID\ procedure is
relatively unaffected by background noise, at the expected level.

	\testbgd

In Figs.~\ref{figure:testexp} we show the effect of variations in data
quality, using exposure times of 15 and $60\ksec$, \ie\/ factors of a
half and twice the standard exposure ({\em TEST-GCF-15K\/} and {\em
TEST-GCF-60K\/}). These models include the cosmic background, and
therefore the reduction in exposure time also results in a degradation
of the signal-to-noise ratio.

	\testexp

In both cases the deconvolved temperature and emissivity profiles form
much better representations of the original data than the convolved
data. (The performance will, of course, differ for clusters with
differing count distributions \ie\/ surface-brightness profiles, and
count rates, \ie\ signal-to-noise for a given exposure.) 

%Despite the
%longer exposure {\em TEST-GCF-60K\/} shows a slight increase in the
%systematic variations of the abundance profile about the original
%data. This is probably due to the systematic effects from the \SID\/
%routine, which will be masked by statistical noise in lower quality
%data. The possible source of this effect is discussed in detail in the
%following section.

	\testnit

\subsection{Systematic Effects From Our Procedure}
\label{section:testnit}
\label{section:syspsf}

In Section~\ref{section:mcarlo} we discussed the Monte-Carlo technique
which we have used to average-out systematic variations which occur in
a single run of the \SID\ procedure. We now discuss some potential
sources for these systematic effects.

The primary assumptions in our procedure are that we can use a
spatially invariant PSF for the image deconvolution phase, and that
the PSF is invariant across each $1\keV$ energy band for the
spectral-reassignment. Before we investigate these issues we will look
at the effect of varying the number of iterations employed in the
image deconvolution phase of the \SID\ procedure.

\subsubsection{The Number of Image Deconvolution Iterations}

In the {\em TEST-NIT-25\/} and {\em TEST-NIT-150\/} tests we vary the
number of iterations employed in the (M-L) image deconvolution stage
of the \SID\/ procedure from the standard number of \itmax, to 25 and
150 iterations, respectively. Figures \ref{figure:testnit}(a) and
\ref{figure:testnit}(b) essentially show that the results are similar
regardless of the number of iterations. A greater number of iterations
will produce sharper features, and thus follow the sharp temperature
decline in the cooling flow models, but this will also exaggerate
spurious features (although this will be minimised by the 10
Monte-Carlo runs of the \SID\ procedure). For this particular cluster
model, we find that \itmax\ iterations appears to be optimal for
recovering of the core temperature decline without introducing severe
systematic effects into the abundance profile.

\subsubsection{The Assumption Of A Spatially Invariant PSF}
\label{section:testpsf}

In Section~\ref{section:mlimage} we described the use of the
Maximum-Likelihood procedure, which requires the use of a spatially
invariant PSF. However, the real \ASCA\/ PSF varies significantly with
position (\eg\/ see Fig.~\ref{figure:psffig}). Thus, we have created
all our simulated data using a spatially varying PSF (interpolating
from the CALDB sample of 11 different PSF positions), and then
deconvolved using a {\em fixed\/} PSF which is appropriate for the
position of the centre of the cluster in the detector image. All the
results presented so far show that this is a reasonable assumption
which does not severely compromise the \SID\/ results. This is because
most of the cluster photons arise from the core region, and therefore
the fixed PSF approximation we have chosen is acceptable.

Our assumption will be poorer for more extended clusters, and for
clusters centred further off-axis because the PSF changes more rapidly
with increasing displacement from the centre of the detector (mainly
in symmetry -- see Fig.~\ref{figure:psffig}). In the {\em
TEST-OFF-30K} test we see the results for a cluster placed near the
edge of the field of view ($9\arcmin$ from the on-axis
position). Although the (abundance profile in particular) results are
worse than for the near on-axis results ({\em TEST-GCF-30K\/}) the
essential trends of the true emissivity and temperature profiles are
recovered. We emphasise that most real cluster observations occur at
the on-axis position we have selected for our standard {\em
TEST-GCF-30K} test.

	\testpsf

\subsubsection{The Assumption Of An Spatially Invariant PSF
Within Each 1 Kev Energyband}

The scatter of the PSF continually degrades with increasing photon
energy. This means that for events at the upper end of the energy band
we will be using a PSF which is too narrow, and for the lower energy
data we will be using a PSF which is too wide. When the
spectral-reassignment procedure determines a map of events between the
{\em source}- and {\em image}-plane it will choose hard events which
are too concentrated, and soft events which are too far away than
would be correct. The precise effect of this is complicated and
probably depends on the spatial distribution of events, but
undoubtedly it will change the slope of the spectral distribution
within each $1\keV$ energy band, resulting in a slight discontinuity
between the boundaries of each energy band. Fortunately, statistical
noise in the data will help obscure these discontinuities, and none
are visible in the deconvolved spectrum of the central region of our
test shown in Fig.~\ref{figure:imgfig}(e).

In summary of all these tests, we believe that our procedure will
provide accurate results under a wide-range of observational
circumstances.

%\subsection{Markevitch \etal's Assumption And Temperature Declines?}
%\label{section:testmm}

%Finally, we investigate the assumption, used by \MFSV, that \ROSAT\/
%data can be used to constrain the cluster emissivity profiles in their
%deconvolution analysis. The question is to whether the \ROSAT\/ data
%($0.2-2\keV$) will be appropriate for the \ASCA\/ data which is over a
%higher energy band ($2-10\keV$ in their analysis). For this test the
%nearest approximation we can make to their assumption is to take the
%deconvolved image from $1-2\keV$ and copy it to the other energy bands
%(rescaling by the number of counts in each band). This effectively
%means that we assume an invariant emissivity profile as a function of
%energy -- as \MFSV\/ have done. The {\em TEST-MM-30K\/}, in
%Fig.~\ref{figure:testmm}, shows that we now obtain a temperature
%decline, similar to the declines seen by \MFSV. Thus, it is possible
%that their assumption, that the emissivity profile in the \ROSAT\/
%energy band can be used for the \ASCA\/ analysis, could be the cause of
%their generic declines.

%	\testmm

%-------------------------

\section{Conclusions}

In this paper we have presented a method for performing a
spectral-imaging deconvolution analysis of X-ray data which has been
affected by an energy- and position-dependent point-spread function
(PSF), such as that produced by the \ASCA\/ satellite mirror
arrangement. Without correction any spatially-resolved spectral
analysis of \ASCA\/ data will be incorrect, as higher energy photons
are scattered further than lower energy photons. In studying spatially
extended sources, such as clusters of galaxies, it is essential that a
correction is made for this effect if accurate conclusion are to be
drawn from the radial variations in the spectral properties of the
object, \eg\/ the temperature and thereby mass profiles of galaxy
clusters.

\citeN{Markevitch:ASCA_A2163_hydeqm} has presented a procedure which 
attempts to correct for the \ASCA\/ PSF, and in an analysis of 30
clusters \citeN{Markevitch:ASCA_Tx_similarity} found that virtually
all their temperature profile results decrease with radius, according
to an average polytropic index of $\gamma=1.24^{+0.20}_{-0.12}$. If
correct, this result is very important as it implies that clusters
have less mass than would be assumed if they were isothermal, and
thereby has implications for cosmology through cluster mass
distributions and baryon fractions. However, the fact that the
polytropic index is so steep, and close to the convective instability
value of $\gamma=5/3=1.67$, is of some concern as many clusters in
their sample are cooling flows, and should therefore be relaxed
clusters. It is therefore imperative that the \MFSV\/ results are
verified by independent means.

We have developed an independent method which is self-contained, in
that it does not require observational data on the same object from
other X-ray satellites, or any assumption about the source
spectrum. This means that potential systematic problems, arising from
cross-calibration uncertainties, are eliminated. Also, unlike
Markevitch \etal, we present tests of our deconvolution procedure on
realistic simulations of \ASCA\/ observations of clusters to show that
our method successfully recovers the true radial properties of the
intracluster gas from data under a wide variety of conditions. In
particular, we have tested the recovery of differing intrinsic
temperature profiles and we have included contamination from the
observed cosmic and instrumental background of \ASCA, to show that the
procedure will work in practical application on reasonably bright
clusters. We have also shown that the various limitations (\ie\/ were
are forced to assume a spatially-invariant point-spread function) do
not unduly compromise the ability of the procedure to successfully
recover the original temperature profiles. All the deconvolved
profiles we obtain are in good agreement with the true intrinsic
properties.

%Through our tests we have shown that assuming a fixed emissivity
%profile leads to incorrect temperature profiles. This test is
%approximately equivalent to the assumption required by
%\citeANP{Markevitch:ASCA_Tx_similarity} in their analysis, as they
%assume that the emissivity profile in the \ROSAT\/ band ($0.2-2\keV$)
%can be used to constrain the emissivity profile required for the
%\ASCA\/ data (above $2\keV$ in their analysis and $1-9\keV$ used in
%this paper). In circumstances where we have forced in an emissivity
%profile which does not vary with energy, we obtain declining
%temperature profiles, similar to those found by
%\citeANP{Markevitch:ASCA_Tx_similarity}

In conclusion, our spectral-imaging deconvolution procedure is
successful at recovering the original radial properties of the
intracluster gas -- for a range of different physical scenarios, and
at the expected level of background and instrumental contamination.
This procedure is, therefore, entirely capable of working in practical
circumstances with observational data. In Paper-II
(\citeNP{White:deconv_ii}) we apply this spectral-imaging
deconvolution procedure to real \ASCA\/ data on \nsample\/ clusters,
and compare the results with those from
\citeANP{Markevitch:ASCA_Tx_similarity} In Paper-III
(\citeNP{White:deconv_iii}) we select a sub-sample of these objects
for mass profile determinations.

%e also propose that an effect
%resulting from one of their assumptions in their methodology may
%xplain the systematic decline in their determination of cluster
%emperature profiles.

%-------------------------

\section{Acknowledgements}

D.A. White acknowledges support from the P.P.A.R.C.; D. Buote that of
P.P.A.R.C. and by NASA through Chandra Fellowship grant PF8-10001
awarded by the Chandra Science Center, which is operated by the
Smithsonian Astrophysical Observatory for NASA under contract
NAS8-39073. We thank S.W. Allen and A.C. Fabian for useful discussion,
and R.M. Johnstone for assistance.

This research has made use of data obtained through the High Energy
Astrophysics Science Archive Research Center Online Service, provided
by the NASA/Goddard Space Flight Center.

The authors thank the anonymous referee for useful comments.

%-------------------------

\bibliography{/data/soft3/astrobibv2.1/bibtex/mnrasmnemonic,/data/daw/text/macros/biblio}
\bibliographystyle{/data/soft3/astrobibv2.1/bibtex/mnrasv2}

%-------------------------

%\cleardoublepage

%\appendix 
%\section{Supplementary Tables}\label{section:appendix}

%	Table~\ref{table:simdat} describes the observational
%	characteristics of the simulation datasets, namely the count
%	statistics of the background-subtracted deconvolved data over
%	the largest region. The background contribution to the more
%	centrally concentrated deconvolution data (compared to the
%	convolved data which also has background contamination),
%	explains why the deconvolved data are extracted from a
%	slightly smaller regions.

%	Table~\ref{table:simresall} presents the resulting spectral
%	fits to the original data (referred to in the `Seq. Num.'
%	column as `org'), the convolved (cnv), and the deconvolved
%	(dnv) data, respectively. (The latter `cnv' and `dnv' results
%	include systematic instrumental effects and background
%	contamination.)

	%\obstab
	%\restab

	%\resfig

%-------------------------

\end{document}